\documentstyle[12pt,epsfig,epsf,feynarts,color]{article}
\def\ga{\mathrel{\raise.3ex\hbox{$>$\kern-.75em\lower1ex\hbox{$\sim$}}}}
\def\la{\mathrel{\raise.3ex\hbox{$<$\kern-.75em\lower1ex\hbox{$\sim$}}}}

\newcommand{\wt}{\widetilde}
\newcommand{\ee}{e^+e^-}

\newcommand{\non}{\nonumber}

\newcommand{\heta}{{\theta}_f}

\def\beq {\begin{equation}}
\def\eeq {\end{equation}}
\def\bec {\begin{center}}
\def\ec {\end{center}}
\def\sq {\widetilde q}
\def\a {\alpha}
\def\b {\beta}
\def\mV {m_{\!V}}

\begin{document}
\begin{flushright}
LPHEA-04-04\\
December 2004 \\
\end{flushright}

\vspace*{0.6cm}
\begin{center}
{\large{ {\bf Third generation sfermions decays 
          into $Z$ and $W$ gauge bosons:
          full one--loop analysis}}}

\vspace{1.5cm}
Abdesslam Arhrib$^{1,2}$ and Rachid Benbrik$^2$ 
\\
1 D\'epartement de Math\'ematiques, Facult\'e des Sciences et Techniques\\
B.P 416 Tanger, Morocco.\\ 
2 LPHEA, D\'epartement de Physique, Facult\'e des Sciences-Semlalia,\\
B.P. 2390 Marrakech, Morocco.\\
\end{center}
$\ $
\\
\\

\begin{abstract}
The complete one--loop radiative corrections to third generation scalar
 fermions into gauge bosons $Z$ and $W^\pm$ is considered. We focus on 
$\wt{f}_2 \to Z \wt{f}_1$ and $\wt{f}_i \to W^\pm \wt{f}_j^\prime$, 
$f,f^\prime =t,b$. We include both SUSY-QCD, QED and full 
electroweak corrections. It is found that the electroweak corrections
can be of the same order as the SUSY-QCD corrections. 
The two sets of corrections
interfere destructively in some region of parameter space.
The full one loop correction can reach 10\% in some SUGRA scenario,
 while in model independent analysis like general MSSM, 
the one loop correction can reach 20\% for large 
$\tan\beta$ and large trilinear soft breaking terms $A_b$.

\end{abstract}
\newpage
\section{Introduction}
\label{sec:1}

Supersymmetric theories predict the  existence of scalar partners 
to all known quarks and
leptons \cite{susy}. In Grand unified SUSY models, the third generation of 
scalar fermions, $\wt{t}, \wt{b}, \wt{\tau}$, gets a special status;
due to the influence of Yukawa-coupling evolution,
the light scalar fermions of the third generation are expected to be lighter 
than the scalar fermions of the first and second generations. 
For the same reason, the splitting between the physical masses
of the third generation may be large enough to allow the opening of the decay 
channels like : $\wt{f}_2 \to \wt{f}_1 V$ and/or $\wt{f}_2 \to \wt{f}_1 \Phi $,
where $V$ is a gauge boson and $\Phi$ is a scalar boson.

Until now there is no direct evidence for SUSY particles,
and under some assumptions on their 
decay rates, one can only set lower limits on their masses \cite{CDF}.
 It is expected that the next generation of $e^+e^-$ machines and/or
hadron colliders (LHC and Tevatron)
could establish the first evidence for the existence of SUSY particles.
Typically, scalar quarks can be produced copiously both at hadron
and lepton colliders. They can in principle be discovered at future
hadron colliders (LHC) up to masses in the
1-2 TeV range while sleptons would become invisible to LHC if heavier
than $\sim 250$ GeV  or so \cite{slep}, due to their weak coupling
and a prominent background. 

If SUSY particles would be detected at hadron colliders,
their properties can be studied with high accuracy at a 
high-energy linear $e^+ e^-$ collider~\cite{LC}. 
It is thus mandatory to incorporate effects beyond leading order
into the theoretical predictions, both for production and decay rate,
in order to match the experimental accuracy.\\
In this spirit,
the next-to-leading order corrections to squark-pair 
production at proton colliders have been studied theoretically 
in \cite{DESY} and found to increase the cross section. 
For $\ee$ machines,
scalar-fermion production has been addressed in 
several studies and shown to be promising for
precision analysis of sfermion
properties with mass and mixing-angle reconstructions~\cite{LC,S.SU}. 
SUSY-QCD corrections to squark-pair production 
at $\ee$ annihilation were shown, a decade ago, to be large 
\cite{ACD,helmut1}.
Recently, the full one-loop radiative corrections to 
the production of scalar muons,
scalar electrons (near threshold) \cite{freitas}, and  third generation
scalar fermions $\wt{t}, \wt{b}, \wt{\tau}$ \cite{helmut2,ah}
have been addressed.
For squark pair production at $e^+e^-$,
the leading and subleading electroweak 
Sudakov logarithms were investigated~\cite{claudio1}
and found to be large at high energy. \\
Similar studies have been carried out for the decays of SUSY particles.
In particular, the QCD corrections to scalar quark decay into quarks plus 
charginos or neutralinos have been studied in \cite{gh1}, while
the full one loop analysis has been addressed in \cite{gh2} and found to 
have important impact on the partial decay widths of scalar fermions.
In Ref.~\cite{DHAJ}, the QCD corrections to the decays of heavy 
scalar quarks into light scalar quarks and Higgs bosons
are found to be of the order $10 \to 20$ \%.

Obviously, most of the studies concentrated  on the 
production and decay of light states $\wt{t}_1, \wt{b}_1$ and 
$\wt{\tau}_1$, while heavier states received less attention 
\cite{ah,gh2,DHAJ,9701336}.
These heavy states can be produced both at LHC and/or at the 
future $e^+e^-$ linear colliders.
The decay of the heavier states third generation scalar fermions
is more complicated than the light one. One can basically have four set
of two-body decays:\\ 
i) Strong decay for stop and sbottom 
$\wt{t}_2\to t \wt{g}$, $\wt{b}_2 \to b \wt{g}$ : 
if these decay are kinematically open they are the dominant one.\\
ii) 
decay to chargino and neutralino : $\wt{f}_2\to f \wt{\chi}_i^0$, 
$\wt{f}_2\to f^\prime \wt{\chi}_i^+$.\\
If the splitting between light and heavy third generation scalar fermions
is large enough we may have the following decays:\\
iii) $\wt{f}_2\to \wt{f}_1 \Phi^0$, 
$\Phi^0=h^0,H^0,A^0$, and $\wt{f}_2\to \wt{f^\prime}_1 H^\pm$.\\
iv) $\wt{f}_2\to \wt{f}_1 Z^0$ and $\wt{f}_2\to \wt{f^\prime}_1 W^\pm$.\\

It has been shown in \cite{qcdv} that the decay modes 
$\wt{f}_2\to \wt{f}_1 Z^0$ and  $\wt{f}_i\to \wt{f^\prime}_j W$, 
if open and under some assumptions, may be the dominant one.
Ref.~\cite{qcdv} also evaluate the gueniun SUSY-QCD corrections
and found them to be of the order -5\% $\to$ -10\%.\\ 
Note also that in several benchmarks scenarios 
for SUSY searches, the bosonic decay of $\wt{t}_i$  and $\wt{b}_i$ 
may be the dominant \cite{SPS}. For example, in SPS5 scenario
the dominant bosonic decay have the following branching ratios \cite{SPSd}:
$Br(\wt{b}_{1}\to W^- \wt{t}_1)=81\%$,
 $Br(\wt{b}_{2}\to W^- \wt{t}_1)=64\%$ and 
$Br(\wt{t}_{2}\to Z^0\wt{t}_1)=61\%$. While in SPS1 scenario,  
we have:  $Br(\wt{b}_{2}\to W^- \wt{t}_1)=34\%$ and 
$Br(\wt{t}_{2}\to Z^0\wt{t}_1)=23\%$.

It is the purpose of this paper to provide the complete one loop radiative 
corrections to $\wt{f}_2\to \wt{f}_1 Z^0$ and  $\wt{f}_i\to \wt{f^\prime}_j W$
including real photon emission, and discuss their effects  
in combination with the SUSY-QCD corrections. \\

The paper is organized as follows. In the next section, 
we will first set the notations and give the  tree-level results.
Section 3 outlines the calculations and  
the on-shell renormalization scheme we will use. 
In section 4, we will discuss the 
effects of radiative corrections
for various types of sfermions decays, 
and we end by  a short conclusion in section 5. 

\section {Notations and  Tree--level formulae}

First we summarize the MSSM parameters needed in our analysis,
with particular attention given to the sfermion sector. 
In the MSSM, the sfermion sector is specified by the mass matrix in 
the basis $(\widetilde{f}_L^{},\widetilde{f}_R^{})$. In terms of 
the scalar mass $\widetilde{M}_L$, $\widetilde{M}_R$, 
the Higgs-Higgsino mass parameter $\mu$ and
the soft SUSY-breaking trilinear coupling $A_f$, the sfermion mass 
matrices squared reads as: 
\begin{equation}
{\cal M}^2_{\widetilde{f}}= 
     \left( \begin{array}{cc} 
                m_f^2 + m_{LL}^2 & m_{LR} m_f \\
                m_{LR} m_f     & m_f^2 + m_{RR}^2
            \end{array} \right) \label{eq1}
\end{equation}
with
\begin{eqnarray}
 m_{LL}^2 &=& \widetilde{M}_{L}^2 
    + m_Z^2\cos 2\beta\,( I_3^f - Q_f s_W^2 ) ,\\
 m_{RR}^2 & = & \widetilde{M}_{R}^2  
                  + m_Z^2 \cos 2\beta\, Q_f s_W^2 , \label{eq:c} \\[2mm]
 m_{LR}    &=& 
 (A_f - \mu\ (\tan\beta)^{-2I_3^f}) \,\, . \label{eq:d}
\end{eqnarray}
$I_3^f=\pm 1/2$ and $Q_f$ are the weak isospin and the electric charge of 
the sfermion $\widetilde{f}$ and $\tan\beta = \frac{v_2}{v_1}$ with 
$v_1$, $v_2$ being the vacuum expectation value of the Higgs fields. 

The hermitian matrix eq.~(\ref{eq1}) is then diagonalized by 
a unitarity matrix 
$R_{{\widetilde{f}}}$, which rotates the current eigenstates,
${\widetilde{f}}_L$ and ${\widetilde{f}}_R$, into the mass
eigenstates $\widetilde{f}_1$ and $\widetilde{f}_2$ as follows:
\begin{equation}
\left( \begin{array}{c} 
                {\widetilde{f}}_1 \\
                {\widetilde{f}}_2
\end{array} \right) = R_{{\widetilde{f}}} 
\left( \begin{array}{c} 
{\widetilde{f}}_L \\
{\widetilde{f}}_R
\end{array} \right) = 
\left( \begin{array}{cc} 
 \cos{{\theta}_f}  & 
 \sin{{\theta}_f}   \\
- \sin{\theta}_f &
 \cos{\theta}_f
            \end{array} \right) 
\left( \begin{array}{c} 
                {\widetilde{f}}_L \\
                {\widetilde{f}}_R
\end{array} \right) \label{eqe}
\end{equation}
where ${\theta}_f$ is the mixing angle such as :
\begin{eqnarray}
\tan 2 {\theta}_f =\frac{2 m_{LR}m_f }{m_{LL}^2 -m_{RR}^2 } \ \ . 
\label{mixing}
\end{eqnarray} 

The mixing angle $\heta$ is proportional to the mass of the fermion
$f$. In the case of the supersymmetric partners of the light fermions, the 
mixing between the
current eigenstates can therefore be neglected.
However, mixing between top squarks can be
sizeable and allows one of the two mass eigenstates
to be much lighter than the top quark. Bottom squark and tau slepton
mixing can also be significant if $\tan\beta$ is large. 

The physical masses, with the convention:
$m_{{\widetilde{f}}_{1}} < m_{{\widetilde{f}}_{2}}$, are given by 
\begin{eqnarray}
m_{{\widetilde{f}}_{1,2}}^2 &=& m_f^2 + \frac{1}{2}(m_{LL}^2 + m_{RR}^2 
\mp \sqrt{ (m_{LL}^2 - m_{RR}^2)^2 + 4 m_{LR}^2 m_f^2 })\ . 
\label{mass} 
\end{eqnarray}

The sfermions sector can be parameterized either by the original 
parameters in the Lagrangian or by the physical masses $m_{\wt{f}_i}=m_i$
and mixing angle $\heta$. Since we are computing radiative corrections
in an on-shell scheme, we will take the following set of physical  parameters:
$$ m_{\wt{t}_2},  m_{\wt{b}_1} m_{\wt{b}_2}, {\theta}_t, {\theta}_b$$  
together with $\mu$ and $\tan\beta$.
Once those parameters are fixed, the light stop mass 
can be derived using the mass sum rule \cite{9701336} as follows:
\begin{eqnarray}
m_{\wt{t}_1}^2=\frac{1}{\cos^2 {\theta}_t}
(  m_W^2\cos 2\beta  -m_{\wt{t}_2}^2 \sin^2 {\theta}_t + 
m_{\wt{b}_2}^2 \sin^2 {\theta}_b + m_{\wt{b}_1}^2 
\cos^2 {\theta}_b +m_t^2 - m_b^2)\label{mass1}
\end{eqnarray}
Of course, eq.~(\ref{mass1}) receive one-loop radiative correction
which we are not included in this analysis. However, it has been shown in 
\cite{thomas2} that the one-loop corrections can shift the tree level mass 
by less than $\la 10$ GeV. 

The soft supersymmetry breaking parameters $A_f$ are then 
connected to the previous ones through:
\beq
A_f=\mu (\tan\beta)^{-2I_f} + \frac{m_{\wt{f}_1}^2 - m_{\wt{f}_2}^2}{m_f} 
\sin {\theta}_f\cos{\theta}_f \label{af}
\eeq
When varying the SUSY parameters, we have to be careful
that charge and color minima do not appear. 
To avoid such minima at tree level, $A_f$ has to satisfy the 
following tree level conditions \cite{ccb} :
\begin{eqnarray}
& & A_t^2 < 3 ( m_{\wt{t}_1}^2 + m_{\wt{t}_2}^2 +
     \frac{1}{2}\cos 2\beta m_Z^2-2 m_t^2 + M_{H_2}^2+ \mu^2   )\nonumber\\
& & A_b^2 < 3 ( m_{\wt{b}_1}^2 + m_{\wt{b}_2}^2 +
     \frac{1}{2}\cos 2\beta m_Z^2-2 m_b^2 + M_{H_1}^2+ \mu^2   )\nonumber\\
& & A_{\tau}^2 < 3 ( m_{\wt{\tau}_1}^2 + m_{\wt{\tau}_2}^2 +
     \frac{1}{2}\cos 2\beta m_Z^2-2 m_{\tau}^2 + M_{H_1}^2+ \mu^2   )
\end{eqnarray}
With $M_{H_2}^2=(m_A^2+m_Z^2)\cos^2\beta-1/2 m_Z^2$ and 
$M_{H_1}^2=(m_A^2+m_Z^2)\sin^2\beta-1/2 m_Z^2$.
For numerical check of CCB as well as $b\to s \gamma $ constraint,
we have used Suspect and Sdecay codes \cite{sdecay1,sdecay2}.

The interaction of the neutral gauge bosons $\gamma$ and $Z$ 
with the sfermion mass eigenstates is described  by the Lagrangian
\begin{eqnarray}
 {\cal L}& =  & 
-i e A^\mu \sum_{i=1,2} Q_f \widetilde{f}_i^* 
\stackrel{\leftrightarrow}{\partial}_\mu
 \widetilde{f}_i - i g_s G_a^\mu \sum_{i=1,2} T^a \widetilde{f}_i^* 
\stackrel{\leftrightarrow}{\partial}_\mu
 \widetilde{f}_i + \nonumber\\
& & i Z^\mu \sum_{i,j=1,2} g_{Z\wt{f}_i \wt{f}_j} 
\widetilde{f}_i^* \stackrel{\leftrightarrow}{\partial}_\mu \widetilde{f}_j
+i W^\mu \sum_{i,j=1,2} g_{W\wt{f}_i \wt{f}^\prime_j} 
\widetilde{f}_i^* \stackrel{\leftrightarrow}{\partial}_\mu 
\widetilde{f^\prime}_j
\label{lag}
\end{eqnarray}
with 
\begin{eqnarray}
& & g_{Z\wt{f}_i \wt{f}_j}= -\frac{e}{s_Wc_W}  \{ 
(I_3^f -Q_f s_W^2) R_{j1}^{\wt{f}} R_{i1}^{\wt{f}} -   
Q_f s_W^2 R_{j2}^{\wt{f}} R_{i2}^{\wt{f}} 
\} \nonumber\\
& & g_{W{\widetilde{f}_i}\widetilde{f^\prime}_{j}} =
-\frac{e}{\sqrt{2}s_{W}}R_{i1}^{f}R_{j1}^{f'}
\end{eqnarray}

The tree--level decay width can thus be written as:
\beq
\Gamma^{0} (\sq^\a_i \to \sq^\b_{j} V) =
\frac{  ( g_{V \wt{f}_i \wt{f}_j} )^2\, \kappa^3 (m_i^2, m_j^2, \mV^2)
}{ 16\pi\, \mV^2\, m_i^3 } ,
\label{tree}
\eeq
with $\kappa (x,y,z) = (x^{2}+y^{2}+z^{2}-2xy-2xz-2yz)^{1/2}$.

\section{Radiative corrections}
\subsection{Scalar fermions decay into gauge bosons at one loop}

\begin{figure}[t!]
\begin{center}
\vspace{-2.1cm}
\input{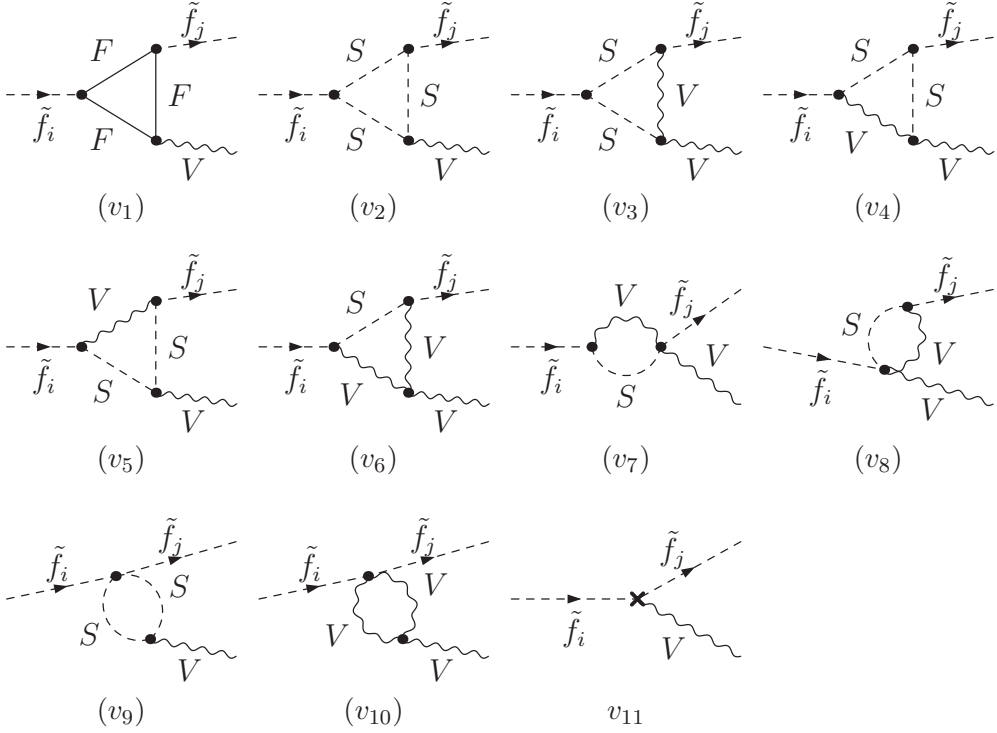}
\vspace{-7cm}
\caption{Generic vertex contributions to 
$\wt{f}_i \to \wt{f}_j^* V$}
\label{vert} 
\end{center}
\end{figure}

\begin{figure}[t!]
\vspace{-2.6cm}
\begin{center}
\input{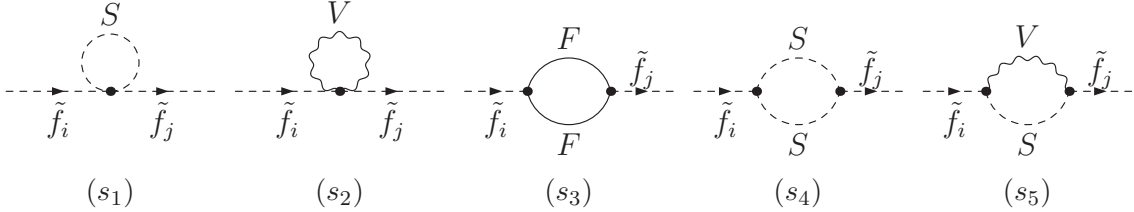}
\vspace{-13cm}
\caption{ Generic Feynman diagrams for 
Scalar fermions self-energies $\wt{f}_i \to \wt{f}_j$ }
\label{self}
\end{center}
\end{figure}

The Feynman diagrams for the one-loop virtual contributions
are generically displayed in
(Fig.~\ref{vert})($v_{1,\ldots,10}$). 
These diagrams are to be supplemented by the external 
self-energy contributions for 
gauge bosons and scalar fermions $\wt{f}_{i,j}$ (Fig.~\ref{self}), 
which are part of the counter-term for vertices 
(Fig.~\ref{vert})($v_{11}$), to be added according to renormalization.
In the generic notation, $V,S,F$ denote all insertions of vector,
scalar, and fermionic states.

At one loop level, transitions between gauge bosons and scalar bosons
like $W^\pm$-$H^\pm$, $W^\pm$-$G^\pm$, $Z^0$-$A^0$, $Z^0$-$G^0$ are present.
Owing to Lorentz invariance, those mixing are proportional to 
$p_V^\mu$ momentum; then since the vector gauge bosons $W$ and $Z$ are 
on-shell transverse, those transitions vanish. In what follows we will
ignore vector-scalar boson mixing.


The full set of Feynman diagrams are generated and evaluated 
using the packages 
FeynArts and FormCalc~\cite{FA}. We have also used
LoopTools and FF~\cite{FF} in the numerical analysis.

We have evaluated the one-loop amplitudes in
the 't Hooft--Feynman gauge. 
The one-loop amplitudes are ultraviolet (UV) 
and infrared (IR) divergent. 
The UV singularities are treated by  dimensional
reduction \cite{siegel}
and are compensated
in the on-shell renormalization scheme. 
We have checked explicitly   
that the results are identical in using dimensional
reduction and dimensional regularization. 
The IR singularities are regularized with a 
small fictitious photon mass $\delta$.

In case of $\wt{f}_2 \to Z \wt{f}_1$ decay,
diagrams like (Fig.~\ref{vert})($v_5$) with $V=\gamma$ or $V=gluon$
and diagram  (Fig.~\ref{vert})($v_{11}$)
are IR divergent. In  (Fig.~\ref{vert})($v_{11}$) the IR divergence 
comes from the wave function renormalization of the scalar fermions.
While for  $\wt{f}_2 \to W \wt{f^\prime}_1$ decay,
diagrams like (Fig.~\ref{vert})($v_{3,...,6}$) and 
(Fig.~\ref{vert})($v_{11}$) $V=\gamma$ or $V=gluon$  are IR divergent,
for an IR-finite decay width we have to add  
the contribution from  real-photon and real-gluon emission,
$\wt{f}_i\to  \wt{f}_j^* V \gamma$ and $\wt{f}_i\to  \wt{f}_j^* V g$ .
\begin{figure}[t!]
\vspace{-4.9cm}
\begin{center}
\input{bremw.tex}
\vspace{-9.5cm}
\caption{Feynman diagrams for real photon (or gluon) emission for in the 
final state of $\wt{f}_i\to  \wt{f}_j V \gamma $ 
(or $\wt{f}_i\to  \wt{f}_j V g $). In case of 
$\wt{f}_i\to  \wt{f}_j' W \gamma $ all diagrams $b_{1,...,5}$ 
contribute, while for $\wt{f}_i\to  \wt{f}_j Z \gamma $ only 
$b_1$, $b_2$ and $b_3$ contribute. In case of QCD brenstrahlung  both for 
$\wt{f}_i\to  \wt{f}_j' W g $  and $\wt{f}_i\to  \wt{f}_j Z g $ only 
$b_1$, $b_2$ and $b_3$ diagrams contribute. }
\label{brem}
\end{center}
\end{figure}

\subsection{Real gluon emission}
In order to cancel the infrared divergence coming from virtual gluon,
the real corrections with an additional gluon in the final state need
also to be included. Feynman diagrams contributing to 
$\delta\Gamma_{g}^{br} = \Gamma(\sq_i \to \sq_j\,V\,g)$  are
drawn in (Fig.\ref{brem}){($b_1$, $b_2$, $b_3$)}.
We would like to mention first, that in the present case and in 
all the following cases, we have checked that the gauge invariance 
is satisfied.\\ 
The three body phase space integration is performed 
following Ref.~\cite{Denner}, which yields a width\footnote{In the 
above eq.~(\ref{glu}), we found that the numerical factor in front of
$I$ integral is 3 in stead of 2 in ref.\cite{qcdv}. 
This disagreement does not affect the numerical result at all.}:
\beq
  \delta\Gamma^{br}_{g} 
  = \frac{g_{V\wt{f}_i\wt{f}_j}^2\,\a_s}{4\pi^2 m_i} \frac{4}{3}\:\left\{ 
  3 I - \frac{\kappa^2}{m_V^2}
  \left[\, I_0 + I_1 + m_i^2\,I_{00} + m_j^2\,I_{11} 
               + (m_i^2 + m_j^2 - m_V^2)\,I_{01} \right] 
  \right\} .\label{glu}
\eeq
where, $\kappa = \kappa(m_i^2,\,m_j^2,\,m_V^2)$, $\alpha_s$ is the
strong coupling constant.
The phase space integrals $I$, $I_{n}$, and $I_{nm}$ 
have $(m_i, m_j, m_V)$ as arguments. 
Their explicit forms are given in \cite{Denner}.
\subsection{Real photon emission}
As in the case of gluon, the infrared divergence coming from virtual photon 
cancels out by including real (soft and hard) photon emission in the
final state.
The diagrams contributing to real bremsstrahlung of 
$\widetilde{f}_i \to \widetilde{f}_j\,Z$ are depicted in 
(Fig.\ref{brem}){($b_1$, $b_2$, $b_3$)}.

In case of $\widetilde{f}_i \to \widetilde{f}_j\,Z\,\gamma$,
the width can be deduced from the gluon bremsstrahlung 
eq.~(\ref{glu}) just by  replacing $\alpha_s$ in eq.~(\ref{glu})
by $\alpha$, eliminating the QCD factor $\frac{4}{3}$ 
and multiplying by the square of scalar fermion charges $e_f^2$.
The width $\delta\Gamma_{\gamma}^{br} = \Gamma(\sq_i \to
\sq_j\,Z\,\gamma)$  
is given by
\begin{eqnarray}
\delta\Gamma^{br}_{\gamma}
& =&\frac{g_{Z\wt{f}_i\wt{f}_j}^2\,\a}{4\pi^2 m_i}e_f^2\:\left\{
  3 I - \frac{\kappa^2}{m_Z^2}
  \left[\, I_0 + I_1 + m_i^2\,I_{00} + m_j^2\,I_{11}
               + (m_i^2 + m_j^2 - m_Z^2)\,I_{01} \right]
  \right\} .\label{phot}
\end{eqnarray}
where $e_d = -\frac{1}{3}$ for down squark and 
$e_{u} = \frac{2}{3}$ for up squark.
\\
Finally, for the Bremsstrahlung $\widetilde{f}_i \to 
\widetilde{f}^\prime_j\, W \gamma$ ,
the Feynman diagrams are drawn  in (Fig.\ref{brem}){$b_{1,\ldots,5}$}.
The decay width is more involving and is given by

\begin{eqnarray}
\delta\Gamma^{br}_{\gamma}
 & =&\frac{g_{W^{\pm}\wt{f}_i\wt{f^\prime}_j}^2\,\a}{4\pi^2 m_i}\:\Big\{
 \frac{3}{4}(e_{f'} + e_f)^2 I -
 \frac{\kappa^2}{m_W^2}\Big[\,\frac{1}{2}e_{f}(e_{f'} + e_{f})I_{0}
 + \frac{1}{2}e_{f'}(e_{f'} + e_{f})I_{1}\nonumber\\
 &+& e_{f}^{2}m_{i}^{2}I_{00}+
 e_{f'}^{2}m_{j}^{2}I_{11}+e_{f}e_{f'}(m_i^2 + m_j^2 -
 m_W^2)I_{01}\Big]\Big\}
\nonumber\\ &+&\frac{\alpha}{8\pi^2 m_{i}}\Big\{3(e_f +
e_{f'})\Big\{g^2_{W^{\pm}\wt{f}_i\wt{f^\prime}_j}(-m^2_j + m^{2}_i)+
g_{G^{\pm}\wt{f}_i\wt{f^\prime}_j}g_{W^{\pm}\wt{f}_i\wt{f^\prime}_j}\Big\}I_{2}\nonumber \\
&-&
\frac{1}{2}\Big\{ g^2_{W^{\pm}\wt{f}_i\wt{f^\prime}_j}(\kappa^2 +
 3m_{W}^2(1-2m_{i}^2-2m_j^2))\nonumber\\ &-&
 6g_{G^{\pm}\wt{f}_i\wt{f^\prime}_j}g_{W^{\pm}\wt{f}_i\wt{f^\prime}_j}(-m_j^2 +
 m_i^2) - 3g^{2}_{G^{\pm}\wt{f}_i\wt{f^\prime}_j}\Big\}I_{22}\nonumber\\
 &-&
\frac{\kappa^2}{m_W^2}e_{f}\Big
\{g^2_{W^{\pm}\wt{f}_i\wt{f^\prime}_j}(m_i^2 - m_j^2 + 2m_{W}^2)+ 
g_{G^{\pm}\wt{f}_i\wt{f^\prime}_j}g_{W^{\pm}\wt{f}_i\wt{f}_j}\Big\}I_{02}
\nonumber\\ &-& \frac{\kappa^2}{m_W^2}e_{f'}\Big
\{g^2_{W^{\pm}\wt{f}_i\wt{f^\prime}_j}( - m_i^2 +
m_j^2 + 2m_{W}^2)+ g_{G^{\pm}\wt{f}_i
\wt{f^\prime}_j}g_{W^{\pm}\wt{f}_i\wt{f^\prime}_j}\Big\}I_{12}\Big\}
\end{eqnarray}
with
\begin{eqnarray}
g_{G^{\pm}\wt{f}_i\wt{f^\prime}_j} &=&\frac{g}{
\sqrt{2}}((m^2_{W}c_{2\b}+ m_f^2 - m_{f'}^2
)R_{i1}^{f}R_{j1}^{f'} +  m_{f'}(\mu \tan_{\b} -
A_{f'})R_{i1}^{f}R_{1j}^{f'}\\\non&+& m_{f}(-\mu
\frac{1}{\tan_{\b}}+A_{f})R_{i2}^{f}R_{1j}^{f'})\\\nonumber
\end{eqnarray}

\subsection{On-shell renormalization}
Recently, there have been several developments
in the renormalization of MSSM. Several schemes 
are available \cite{thomas2,itp,vienna,thomas1}.
Here, we follow the strategy of~\cite{gh2} by introducing 
counter-terms for the physical parameters, i.e.\ for masses and
mixing angles, and perform  field renormalization
in a way that residues of renormalized propagators 
can be kept at unity.

We will adopt throughout, the on--shell renormalization scheme of 
Refs. \cite{Denner} for SM parameters and fields. 
We make the following prescriptions:
\begin{eqnarray}
& & e \to (1+\delta Z_e) e \qquad , \qquad M_{W,Z} \to M_{W,Z} +
\delta M_{W,Z}\label{cd1}
\end{eqnarray}
the gauge bosons are renormalized such as :
\begin{eqnarray}
Z^\mu \rightarrow Z_{ZZ}^{1/2} Z^\mu+
Z_{Z\gamma}^{1/2} A^\mu \ \ , \ 
A^\mu \rightarrow Z_{AA}^{1/2} A^\mu + Z_{\gamma Z}^{1/2} Z^\mu \ \ , \ \ 
W^\mu \rightarrow (1+ \frac{1}{2} \delta Z_{WW}^{1/2}) W^\mu 
 \label{cd2}
\end{eqnarray}
with $Z_{ij}^{1/2} = \delta_{ij} + \frac{1}{2} \delta Z_{ij} $.
In the on--shell scheme we use the mixing angle $s_W$ (resp $c_W$) 
is defined by 
$s_W^2=1-M_W^2/M_Z^2$ (resp $c_W^2=M_W^2/M_Z^2$). Its counter--term 
is completely fixed by the mass counter-terms of W and Z gauge bosons as:
\begin{eqnarray}
\frac{\delta s_W}{s_W}= -\frac{1}{2} \frac{c_W^2}{s_W^2} \Bigm ( 
\frac{\delta M_W^2}{M_W^2} - \frac{ \delta M_Z^2}{M_Z^2} \Bigm)\quad , \quad
\frac{\delta c_W}{c_W}= -\frac{1}{2} \Bigm ( 
\frac{\delta M_W^2}{M_W^2} - \frac{\delta M_Z^2}{M_Z^2} \Bigm)
\end{eqnarray}

The extra parameters and fields we still have to renormalize in our case
are the scalar fermion wave functions $\wt{f}_i$ and the mixing angle
${\theta}_{f}$ defined in eq.~(\ref{mixing}).

In the general case, where sfermions mixing is allowed, 
the wave functions of the two
sfermions mass eigenstates are not decoupled. 
Taking into account the mixing, the renormalization of the
sfermions wave functions and the mixing angle 
$\heta$ can be performed by making the
following substitutions in the Lagrangian eq.~(\ref{lag})
\begin{eqnarray}
\widetilde{f}_1 \rightarrow Z_{11}^{1/2} \wt{f}_1+
Z_{12}^{1/2} \wt{f}_2 \ \ , \ 
\wt{f}_2 \rightarrow Z_{22}^{1/2} \wt{f}_2 + Z_{21}^{1/2} \wt{f}_1 \ \
, \ \ 
\heta \rightarrow \heta+\delta \heta  \label{cd3}
\end{eqnarray}

Using the above prescriptions Eq. (\ref{cd1}-\ref{cd2}, \ref{cd3}) 
in the Lagrangian (\ref{lag}),
the Lagrangian counter terms can be obtained and is given by
\begin{eqnarray}
 \delta {\cal L} & = & \sum_{i,j=1,2} [  
 \delta(Z \wt{f}_i\wt{f}_j) Z^\mu 
\widetilde{f}_i^* 
\stackrel{\leftrightarrow}{\partial}_\mu \widetilde{f}_j +
\delta(W \wt{f}_i\wt{f^\prime}_j) W^\mu 
\widetilde{f}_i^* \stackrel{\leftrightarrow}{\partial}_\mu 
\widetilde{f^\prime}_j]
\label{dlag}
\end{eqnarray}
where
\begin{eqnarray}
 \delta(Z \wt{f}_i\wt{f}_j) & = & -e Q_f \frac{1}{2} \delta Z_{\gamma Z} 
+ g_{Z\wt{f}_i \wt{f}_j} (\delta Z_e + \frac{1}{2} \delta Z_{ZZ}  +
\frac{1}{2} \delta Z_{ii} + \frac{1}{2} \delta Z_{jj})
\nonumber\\ &&
+ \frac{\delta s_W e }{c_W^3 s_W^2}
  ((-I_3^f - Q_f s_W^2 + 2 I_3^f s_W^2) R_{j1}^{\wt{f}} R_{i1}^{\wt{f}}
- Q_f s_W^2 R_{j2}^{\wt{f}} R_{i2}^{\wt{f}}  )
\nonumber \\ & & +
g_{Z\wt{f}_k \wt{f}_j}\delta Z_{ki} +g_{Z\wt{f}_i \wt{f}_l}
\delta Z_{lj} 
+ \Delta(g_{Z\wt{f}_i \wt{f}_j}) \delta \heta\nonumber\\
\delta(W{\widetilde{f}_i}\widetilde{f}_{j}^{'}) & = &
g_{W{\widetilde{f}_i}\widetilde{f}_{j}^{'}}(\frac{\delta Z_{ii}}{2} +
\frac{\delta Z_{jj}}{2} )+g_{W{\widetilde{f}_i}\widetilde{f}_{l}^{'}}
\frac{\delta Z_{lj}}{2}+g_{W{\widetilde{f}_k}\widetilde{f}_{j}^{'}}
\frac{\delta Z_{ki}}{2} + \non \\
&&  g_{W{\widetilde{f}_i}
\widetilde{f}_{j}^{'}} (\frac{\delta Z_{WW}}{2} + \frac{\delta s_{W}}{s_W} +
\delta Z_{e}) - \Delta(g_{W{\widetilde{f}_i}
\widetilde{f}_{j}^{'}})\Bigm) 
\end{eqnarray}
where 
\begin{eqnarray}
\Delta(g_{Z\wt{f}_1 \wt{f}_1})&=&-\Delta(g_{Z\wt{f}_2 \wt{f}_2})=2
g_{Z\wt{f}_1 \wt{f}_2}\ \ , \ \ \Delta(g_{Z\wt{f}_1
  \wt{f}_2})=\Delta(g_{Z\wt{f}_2 \wt{f}_1})= g_{Z\wt{f}_2
  \wt{f}_2}-g_{Z\wt{f}_1 \wt{f}_1} \nonumber\\
\Delta(g_{W{\widetilde{f}_1}\widetilde{f}_{1}^{'}})
& =& g_{W{\widetilde{f}_2}\widetilde{f}_{1}^{'}} 
\delta\theta_{f} + g_{W{\widetilde{f}_1}\widetilde{f}_{2}^{'}} 
\delta\theta_{ f^{'}}\quad , \quad
\Delta(g_{W{\widetilde{f}_2}\widetilde{f}_{2}^{'}})
= -g_{W{\widetilde{f}_1}\widetilde{f}_{2}^{'}} 
\delta\theta_{f} - 
g_{W{\widetilde{f}_2}\widetilde{f}_{1}^{'}} \delta\theta_{f^{'}}
\non\\
\Delta(g_{W{\widetilde{f}_1}\widetilde{f}_{2}^{'}})
& =& g_{W{\widetilde{f}_2}\widetilde{f}_{2}^{'}} 
\delta\theta_{f} - 
g_{W{\widetilde{f}_1}\widetilde{f}_{2}^{'}} \delta\theta_{f^{'}}
\quad , \quad 
\Delta(g_{W{\widetilde{f}_2}\widetilde{f}_{1}^{'}})
 = -g_{W{\widetilde{f}_1}\widetilde{f}_{1}^{'}} 
\delta\theta_{f} 
+ g_{W{\widetilde{f}_2}\widetilde{f}_{2}^{'}} 
\delta\theta_{f^{'}}
\end{eqnarray}

To fix all the above renormalization constants,
we use the following renormalization conditions:

\begin{itemize}
\item The on-shell conditions for $m_W$, $m_Z$, $m_e$ 
and the electric charge $e$ are defined as in the Standard 
Model \cite{Denner}.

\item On-shell condition for the scalar fermion $\widetilde{f}_i$ :
we choose to identify
the physical scalar fermion mass with the corresponding parameter in
the renormalized Lagrangian,
and require the residue of the propagators to have its tree-level 
value, i.e., 
\begin{eqnarray}
\delta Z_{ii} & = & -\Re\{\frac{\partial}{\partial p^2}(
{\Sigma}_{\wt{f}_i \wt{f}_i} (p^2))\} |_{p^2=m^2_{\wt{f}_i} }   
 \ \ , \
\   
\delta Z_{ij}=
\frac{\Re\{{{\Sigma}}_{ \wt{f}_i \wt{f}_j }(m_{\wt{f}_j}^2)\}}
{m_{\wt{f}_j}^2-m_{\wt{f}_i}^2} \quad , \quad \nonumber \\
\delta
m^2_{\wt{f}_i}  & = &  \Re ({\Sigma}_{\wt{f}_i \wt{f}_j}
(m^2_{\widetilde{f}_i}))
\end{eqnarray}
where $\sum_{\wt{f}_i \wt{f}_j } (p^2)$, $i,j=1,2$ is the scalar 
fermion bare self-energy.
\end{itemize} 

One has then to choose a renormalization condition which defines the mixing
angle $\heta$. We select this condition in such a way to kill the 
transitions $\wt{f}_i \leftrightarrow \wt{f}_j$ at the one--loop level.
The renormalization of the scalar fermion mixing 
angle is then given by \cite{gh2}:
\begin{eqnarray}
\delta{\theta}_{{f}} = \frac{1}{2} \frac{\Sigma_{\wt{f}_i \wt{f}_j}(
m^2_{ \wt{f}_j} )+ \Sigma_{\wt{f}_i \wt{f}_j}(m^2_{ \wt{f}_i} ) }
{ m^2_{ \wt{f}_j} - m^2_{ \wt{f}_i} } \label{angle}
\end{eqnarray}

\section{Numerics}
Now we are ready to 
present our numerical results both for the tree-level and one-loop 
decay widths and branching ratios for $\tilde{f}_i \to \tilde{f}_j Z$ and
$\tilde{f}_i \to \tilde{f'}_j W^{\pm}$. Let us first fix our
inputs and SUSY parameters choice.

As experimental data points~\cite{pdg},
the following input quantities enter:
$\alpha^{-1}=137.03598$, $m_Z=91.1875$ GeV, 
$m_W=80.45$ GeV. Fermions masses are given by:
\begin{eqnarray}
&&  m_e= 0.000511\ \ {\rm GeV} \,, m_{\mu} = 0.1056\ \ {\rm GeV} \,, 
m_{\tau} = 1.777\ \ {\rm GeV} \,, \nonumber \\
& & m_t = 178\ \ {\rm GeV} \,,  m_b = 4.7\ \ {\rm GeV} \,, 
m_c = 1.5\ \ {\rm GeV} \,, 
m_u = 0.062\ \ {\rm GeV} \,, \nonumber\\ &&  m_d = 0.083\ \
{\rm GeV}  \,,  m_s = 0.215\ \ {\rm GeV} \, \nonumber
\end{eqnarray}
where effective quark masses reproducing the hadronic vacuum polarization
contribution $\Delta \alpha(m_Z^2)$ with a sufficiently high accuracy
have been chosen  \cite{Eidelman}.

For the SUSY parameters, we will use MSSM inputs which look like
some of  the Snow-mass Points and Slopes (SPS) and 
benchmarks scenarios for SUSY searches \cite{SPS}.
For our study we will use SPS1 and SPS5 scenario. As we explained
in the introduction, for those 2 scenarios the bosonic decays 
of scalar fermions $\wt{f}_i \to  \wt{f}_j V$, when open, are dominant.
More details about the mass spectrum and decays rates can be found in 
\cite{SPS,SPSd}. 
\begin{figure}[t!]
\smallskip\smallskip 
\vskip-1.3cm
\centerline{
{\epsfxsize3.6 in\epsffile{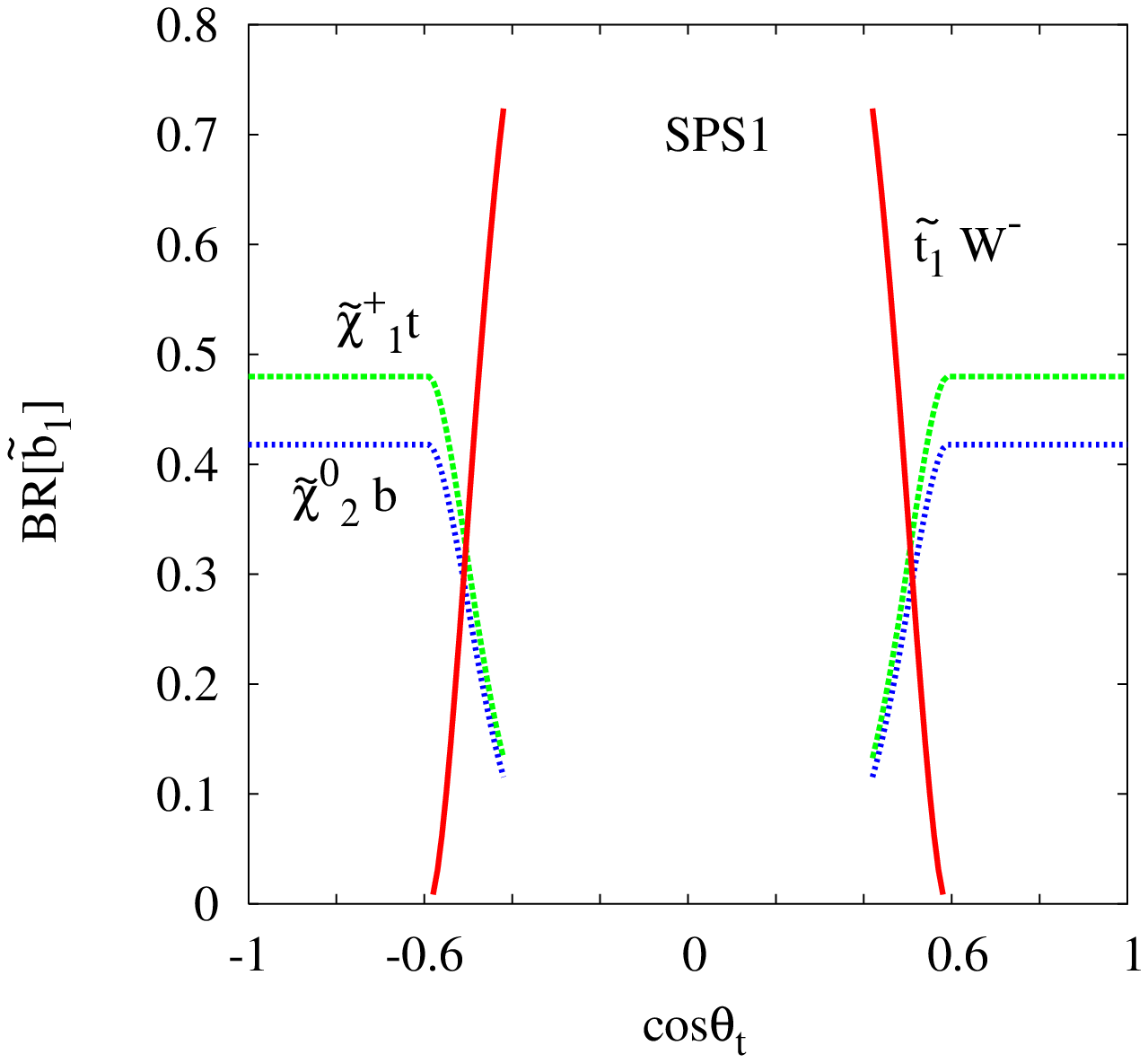}}  
\hskip-1.99cm
{\epsfxsize3.61 in\epsffile{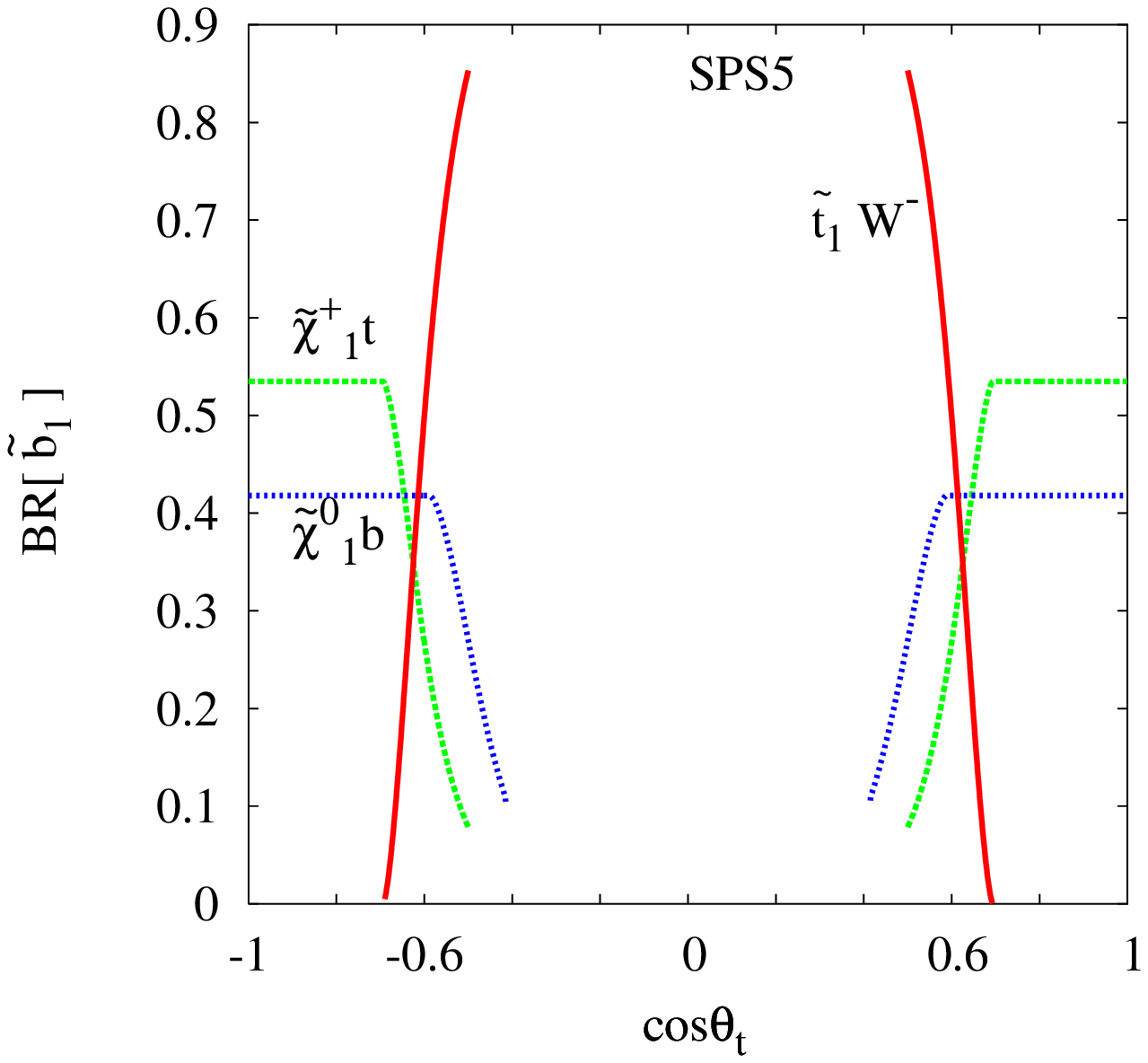}}}
\hskip-1.4cm
\smallskip\smallskip
\vskip-1.5cm
\centerline{
{\epsfxsize3.6 in\epsffile{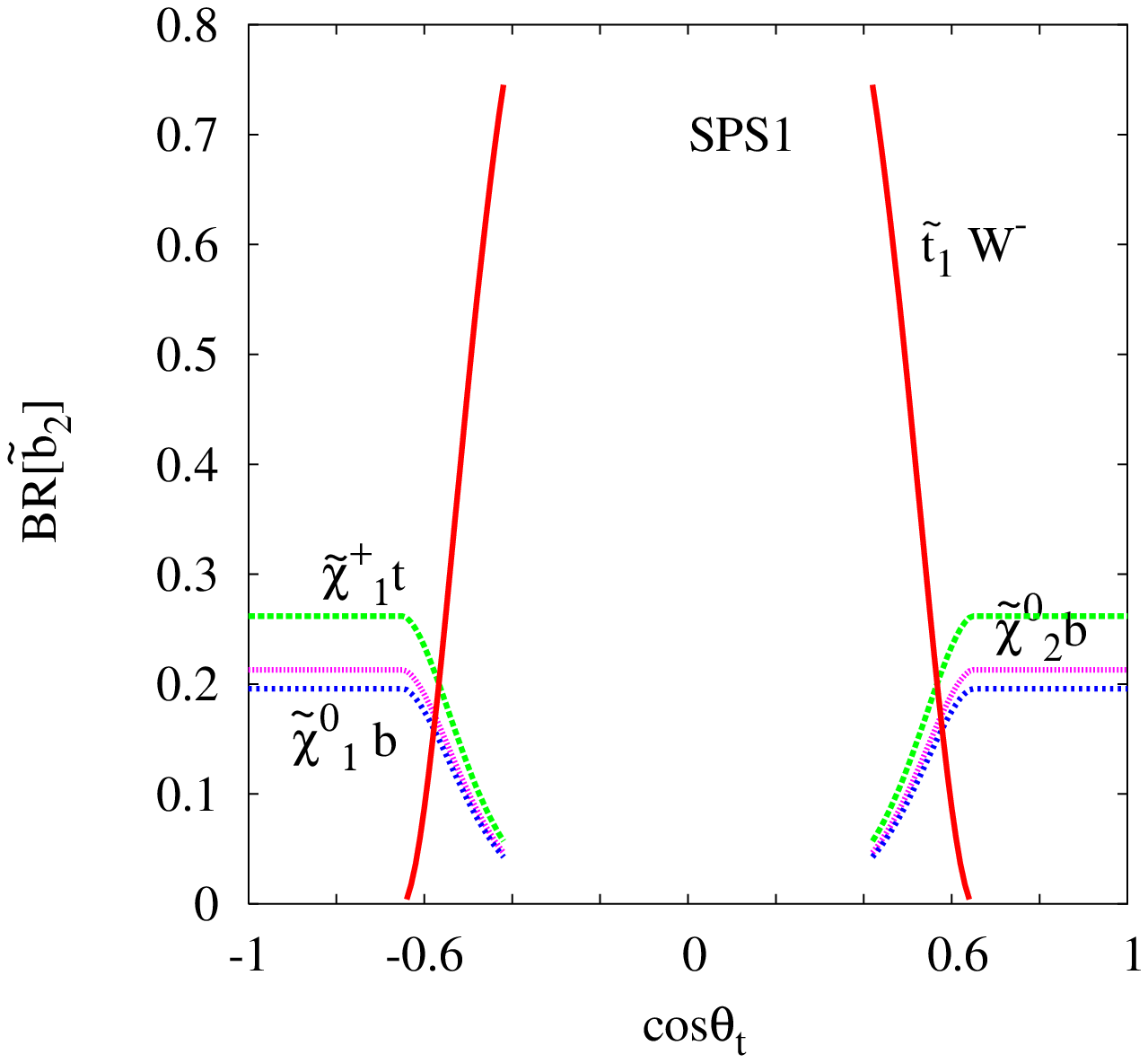}}  
\hskip-1.99cm
{\epsfxsize3.61 in\epsffile{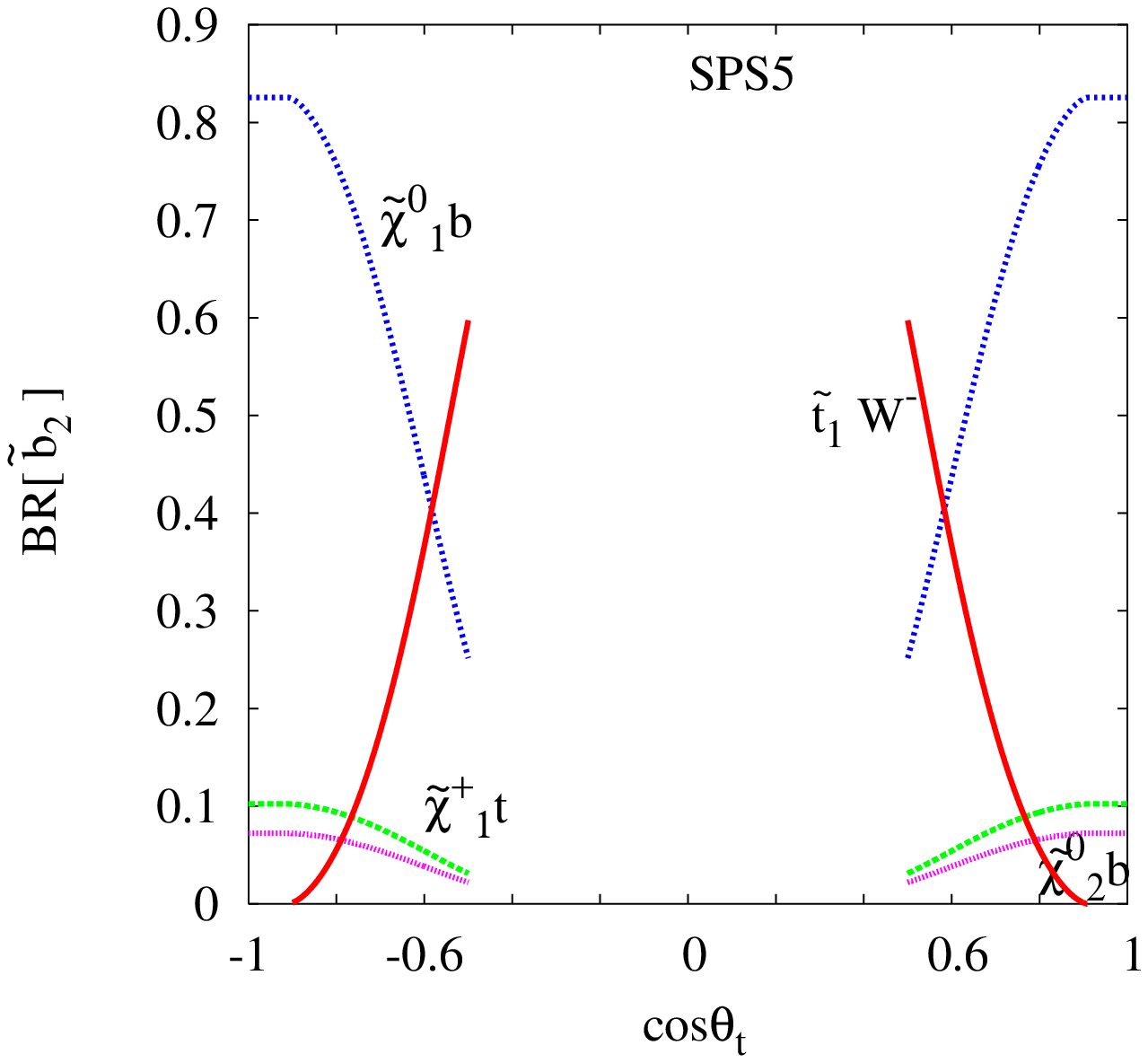}}}
\hskip-1.4cm
\smallskip\smallskip
\vskip-1.5cm
\centerline{
{\epsfxsize3.6 in\epsffile{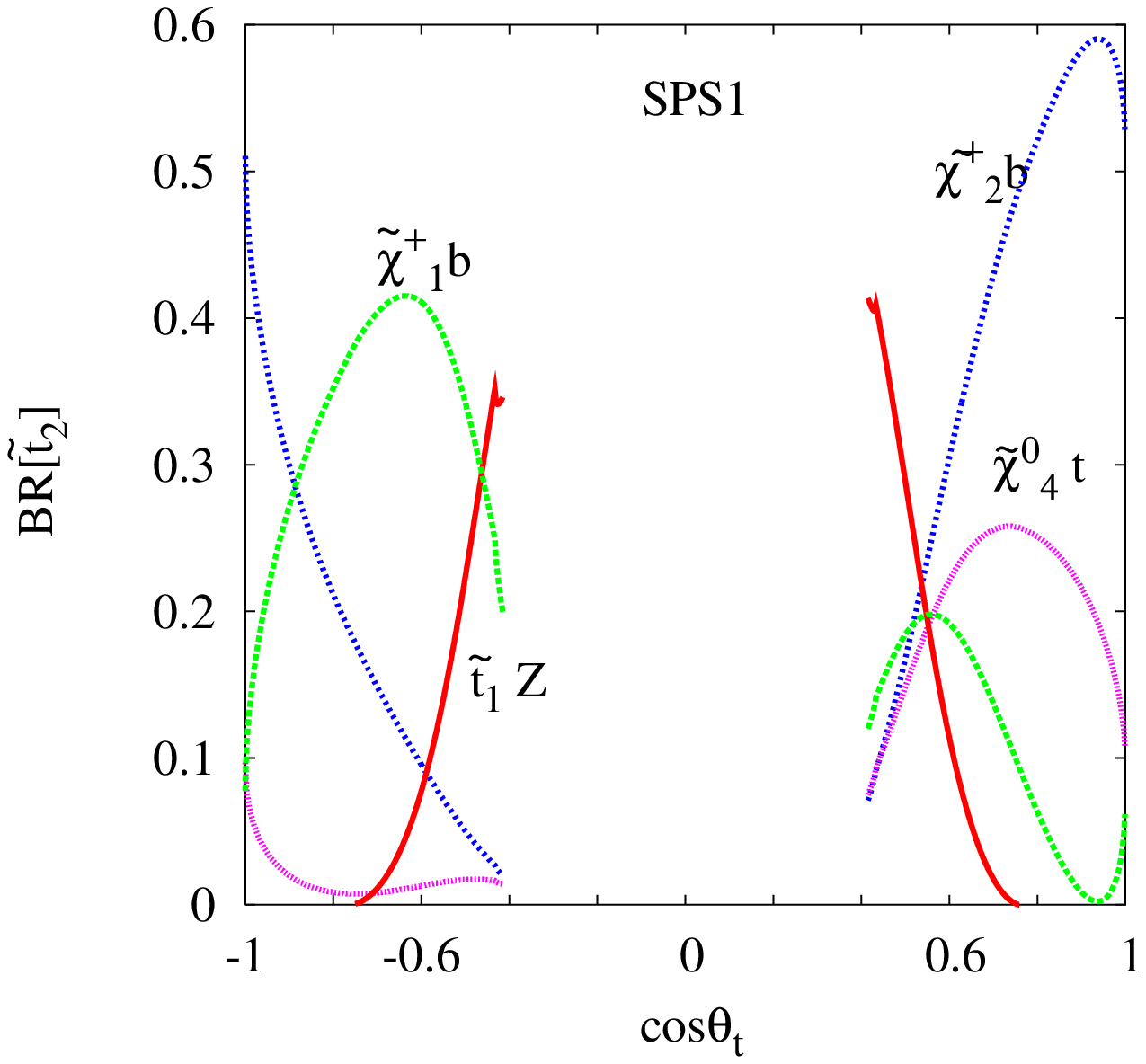}}  
\hskip-1.99cm
{\epsfxsize3.61 in\epsffile{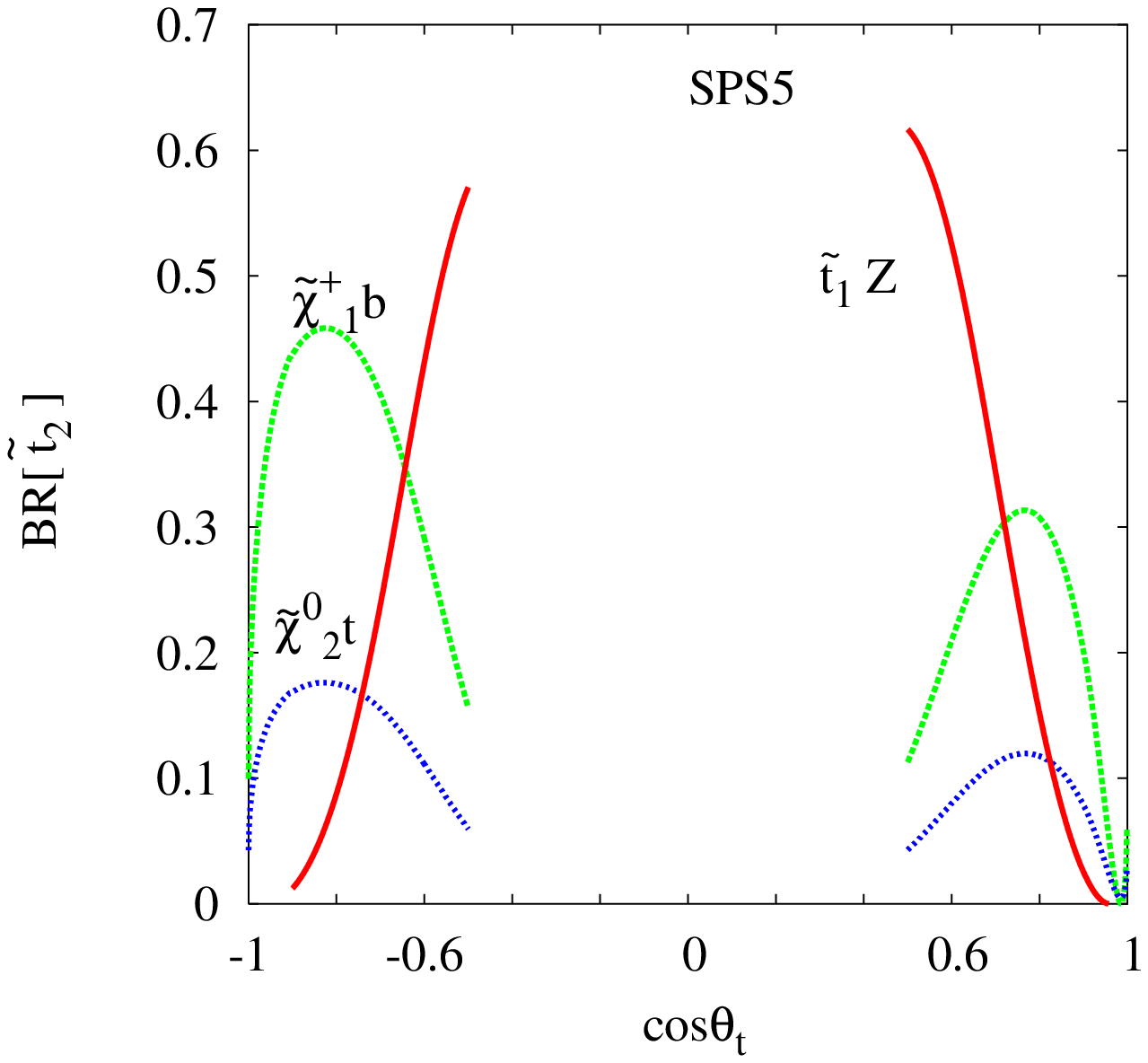}}}
\smallskip\smallskip
\vskip-.6cm
\caption{Branching ratios of bosonic decays of $\wt{b}_1$ (upper
  plots), $\wt{b}_2$ (middle plots)
and $\wt{t}_2$ (lower plots) in SPS1 (left) and SPS5 (right) as 
function of $\cos\theta_t$.  }
\label{fig4}
\end{figure}

In SPS1, we have the following spectrum (are listed only the
parameters needed):
 $\tan\beta=10$, $m_{A^0} = 394$ GeV, 
$A_t = -431.34$ GeV, $A_b =582.67 $ GeV, 
$M=193$ GeV, $M'=99$ GeV, $\mu=352$ GeV. The mass of the first and 
second generation scalar fermion is of the order 177 GeV (average).
While the masses of the third generation scalar fermions are :
$m_{\wt{t}_{1}} = 396.43$ GeV, $m_{\wt{t}_{2}} = 574.71$ GeV ,
$m_{\wt{b}_{1}} = 491.91$ GeV ,  $m_{\wt{b}_{2}} = 524.59$ GeV.
The mixing angle are given by 
$\cos\theta_t = 0.57$ , $\cos\theta_b = 0.88$.


In SPS5 (light stop scenario), we have the following spectrum :
$m_{A^0} =  694$ GeV, $\tan\beta=5$
$A_t = -785.57$ GeV, $A_b = -139.11$ GeV, 
$M=235$ GeV, $M'=121$ GeV, $\mu= 640 $ GeV
The mass of the first and second generation scalar 
fermion is of the order 231 GeV (average).
The masses of the  third generation are
$m_{\tilde t_{1}} = 253.66$ GeV , $m_{\tilde t_{2}} = 644.65$ GeV ,
$m_{\tilde b_{1}} = 535.86$ GeV ,
$m_{\tilde b_{2}} = 622.99$ GeV.
The mixing angle are given by
$\cos\theta_t = 0.54$ , $\cos\theta_b = 0.98$.

In fact, our strategy is the following :
the SPS1 and SPS5 outputs are fixed as above,
but we will allow a variation of the mixing angles $\cos\theta_t$, 
$\cos\theta_b$ from their SPS values.
According to our parametrization defined in section 2,
we choose as independent parameters
$m_{\wt{t}_2},  m_{\wt{b}_1} m_{\wt{b}_2}, {\theta}_t, {\theta}_b$ 
together with $\mu$ and $\tan\beta$. $m_{\wt{t}_1}$ is fixed by 
eq.~(\ref{mass1}) and the soft trilinear parameters 
are fixed using eq.~(\ref{af}). 
The variation of $\cos\theta_t$
and $\cos\theta_b$ imply the variation of $m_{\wt{t}_2}$ 
as well as $A_t$ and  $A_b$.
Since we allow  variation of the
$\cos\theta_t$ and $m_{\wt{t}_2}$ mass, our inputs 
can be viewed as a general MSSM inputs and not as SPS one.\\ 
As outlined in section~2, $A_{t,b}$ are fixed by tree level relation  
eq.~(\ref{af}). Of course, $A_{t,b}$ receive radiative corrections 
at high order. However, $A_{t}$ and $A_b$ enter game only at one-loop level
in our processes, radiative corrections to $A_{t}$ and $A_b$ is considered
as two-loop effects.  \\
Before presenting our results, we would like to mention that we 
neglect radiative corrections to eq.~(\ref{mass1}). 
As mentioned in section 2, the one-loop effect correction to eq.~(\ref{mass1})
can shift the tree level masses only by less than 
$\la 10$ GeV. We have checked that for our process
such shift does not affect significantly our result.

In Fig.~(\ref{fig4}) we show branching ratios of 
$\wt{b}_{1}$, $\wt{b}_{2}$ and $\wt{t}_{2}$.
We evaluate the bosonic decays : 
$\wt{b}_{1}\to W^- \wt{t}_1$,
$\wt{b}_{2}\to W^- \wt{t}_1$ and 
$\wt{t}_{2}\to Z^0\wt{t}_1$ 
as well as the fermionic decays  
$\wt{f}_{i}\to \chi_i^{0} f (\chi_i^{+} {f^\prime})$
as function of $\cos\theta_t$ for SPS1 (left) and SPS5 (right) scenario.
From those plots, it is clear that the bosonic decay,
once open, are the dominant
one for $|\cos\theta_t|\approx 0.4\to 0.45$. For $|\cos\theta_t|\approx 0.4$ 
the light stop $m_{\wt{t}_1}$ is about $100$ GeV, when
$|\cos\theta_t|$ 
increases, the  $m_{\wt{t}_1}$ increases also and for large $|\cos\theta_t|$ 
the bosonic decays are already close and the branching ratio vanishes.

We note that in the case of SPS1 
the bosonic decays are open only for $0.4\la |\cos\theta_t| \la 0.6$ 
Fig~.(\ref{fig4}) (left). 
In the region $|\cos\theta_t| \la 0.4$, the light stop is below
the experimental upper limit $m_{\wt{t}_1}\approx 90 $ GeV, and no data
are shown. While in the case of SPS5 Fig~.(\ref{fig4}) (right), for 
$|\cos\theta_t| \la 0.5$,
we find that $m_{\wt{t}_1}$ is below the experimental upper limit
and also $\delta\rho \ga 0.001$ due to large splitting
between stops and sbottoms.


The magnitude of SUSY radiative corrections can be described by
the relative correction which we define as:
\begin{eqnarray}
\Delta = \frac{\Gamma^{\rm{1-loop}}(\wt{f}_i \to \wt{f}_j V)-
\Gamma^{\rm{tree}}(\wt{f}_i \to \wt{f}_j V)}
{\Gamma^{\rm{tree}}(\wt{f}_i \to \wt{f}_j V)}\label{del}
\end{eqnarray}
\begin{figure}[t!]
\smallskip\smallskip 
\vskip-1.5cm
\centerline{{
\epsfxsize3.7 in 
\epsffile{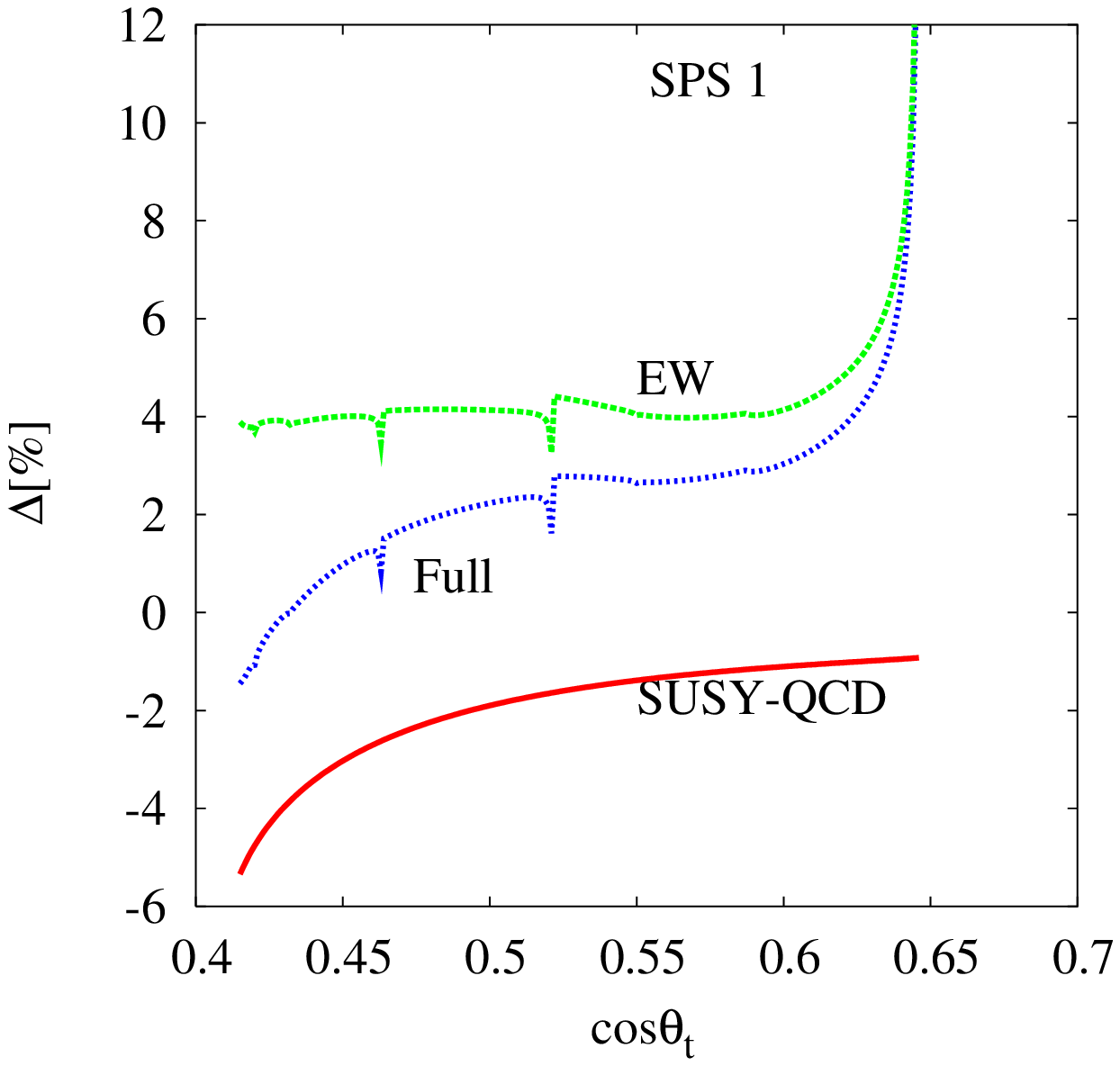}}  \hskip-1.99cm
\epsfxsize3.7 in 
\epsffile{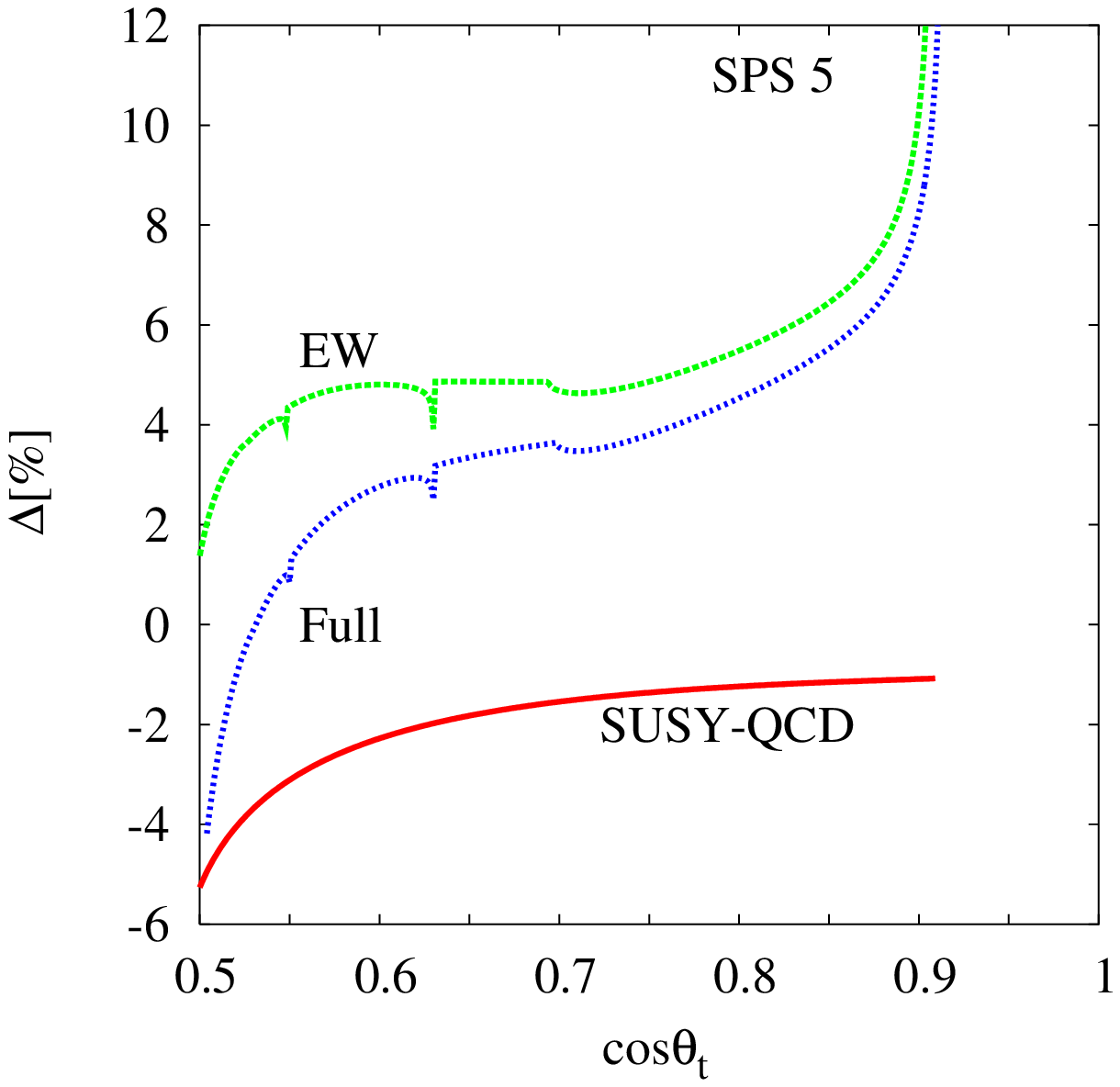} }
\smallskip\smallskip
\caption{Relative correction (electroweak EW, SUSY-QCD and full) 
to $\wt{b}_2 \to \wt{t}_1 W$ 
as function of $\cos\theta_t$ in SPS1 (left) and SPS5 (right)}
\label{fig5}
\end{figure}
\begin{figure}[t!]
\smallskip\smallskip 
\vskip-1.cm
\centerline{
{\epsfxsize3.7 in\epsffile{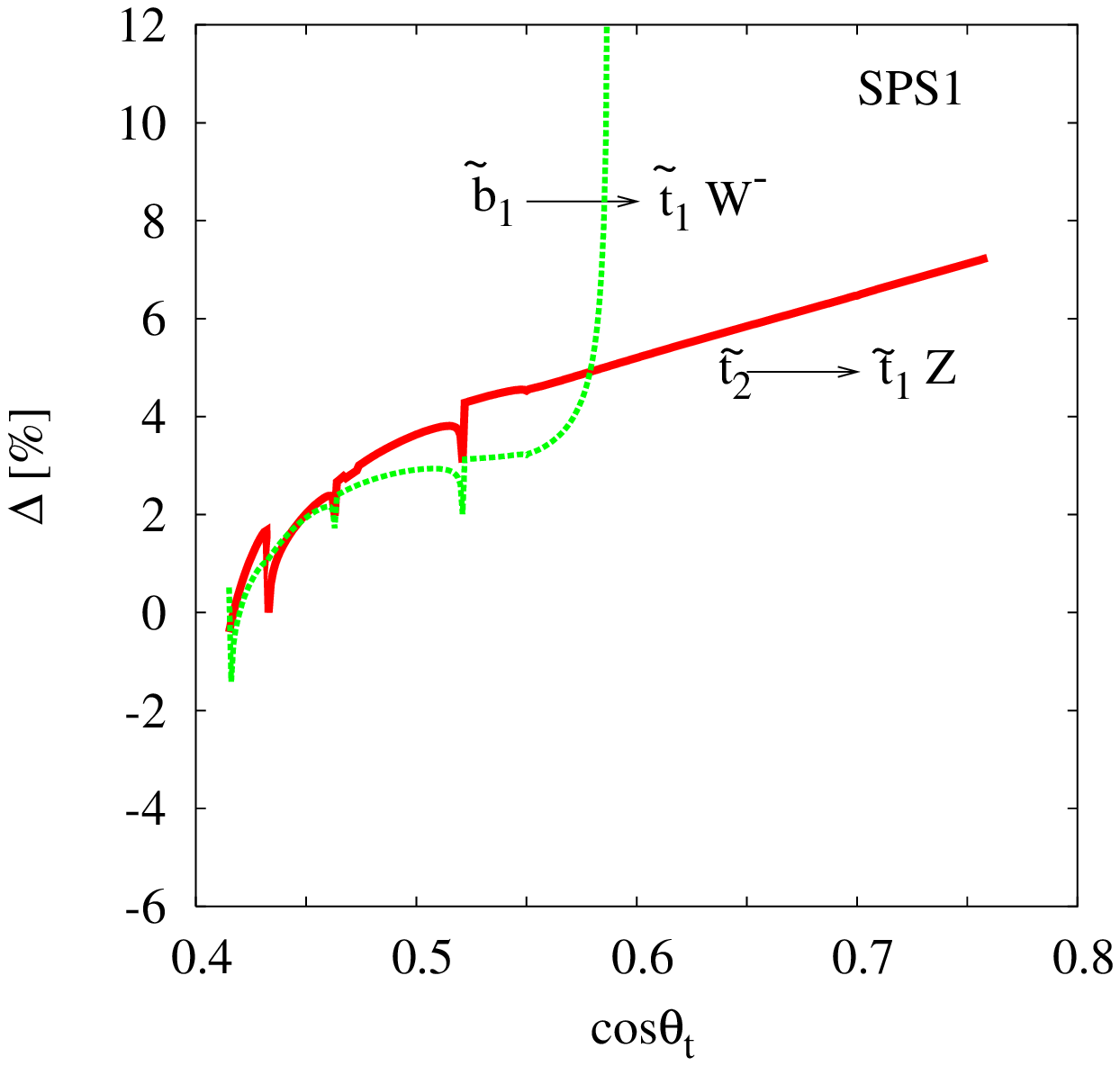}}  
\hskip-1.99cm
{\epsfxsize3.7 in\epsffile{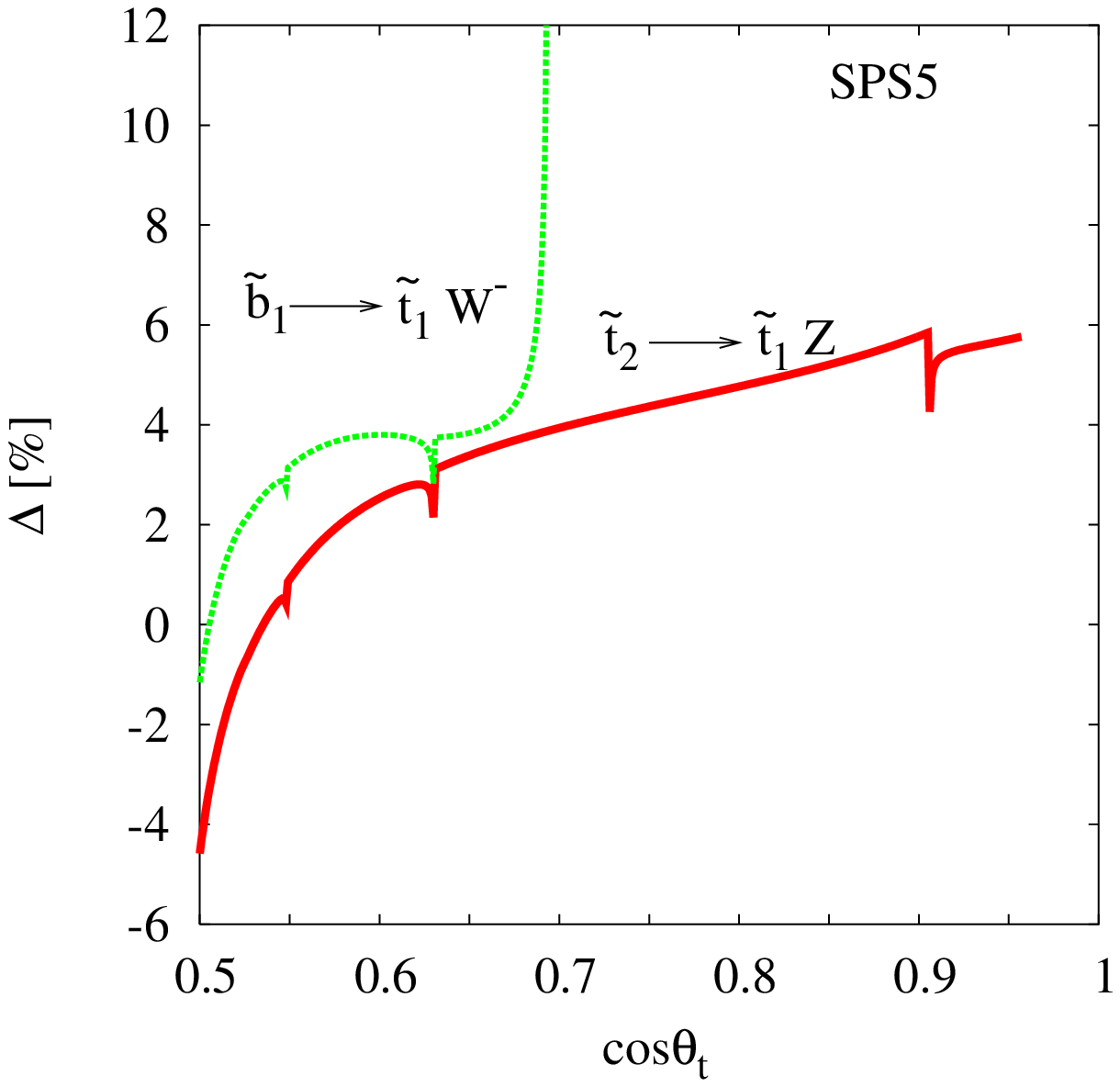}}}
\smallskip\smallskip
\vskip-.40cm
\caption{Relative correction to $\wt{b}_1 \to \wt{t}_1 W$ and 
 $\wt{t}_2 \to \wt{t}_1 Z$ 
as function of $\cos\theta_t$ in SPS1 (left) and SPS5 (right)}
\label{fig55}
\end{figure}

In Fig.~(\ref{fig5}) we illustrate the relative correction
$\Delta$ as function of $\cos\theta_t$ for the decay
$\wt{b}_2 \to W \wt{t}_1$ in SPS1 (left) and SPS5 (right).
As it can be seen from the left plot, the SUSY-QCD corrections lies in
the range $-1\% \to -6\%$ while the EW corrections lie in the range
$4\% \to 10\%$ for $\cos\theta_t \approx 0.4 \to 0.65$.
The SUSY-QCD and EW corrections are of opposite sign,
there is a destructive interference and so the full one-loop 
corrections lie between them.
For $\cos\theta_t\to 0.65$,
the total correction increases to about 10\%. 
This is due to the fact that for $\cos\theta_t\to 0.65$
the mass of light stop is $m_{\wt{t}_1}\approx 444$ GeV, the decay 
$\wt{b}_2 \to W \wt{t}_1$ is closed and so the tree level 
width decreases to zero. 
The observed peaks around $\cos\theta_t\approx 0.46$ 
(resp $\cos\theta_t\approx 0.53$)
correspond to the opening of the transition $\wt{t}_1\to \chi_1^0 t$
(resp $\wt{t}_1\to \chi_2^0 t$).
The right plot of Fig.~(\ref{fig5}) in SPS5 scenario, 
exhibits almost the same behavior as the left plot. 
The electroweak corrections interfere destructively with the SUSY-QCD
ones, the full corrections
are between $-4\% \to 10\%$ for $\cos \theta_t\in[0.5,0.9]$.

In Fig.~(\ref{fig55}) we show the relative correction
$\Delta$ as function of $\cos\theta_t$ for the decay
$\wt{b}_1 \to W \wt{t}_1$ and  
$\wt{t}_2 \to Z \wt{t}_1$  in SPS1 (left) 
 and SPS5 (right) scenario.\\
In the case of $\wt{t}_2 \to Z \wt{t}_1$,
the total correction lies in $-1\to  7\%$ (resp $-5 \to 6\%$) 
in SPS1 (resp SPS5) scenario. 
From Fig.~(\ref{fig55}), one can see that 
the relative corrections for $\wt{b}_1 \to W \wt{t}_1$
are enhanced for $\cos\theta_t\approx 0.6$ (resp $\cos\theta_t\approx
0.75$) in SPS1 (resp SPS5). This behavior has the same explanation as 
for $\wt{b}_2 \to W \wt{t}_1$ in figure.~(\ref{fig5}).
At $\cos\theta_t\approx 0.6$ (resp $\cos\theta_t\approx
0.75$) in SPS1 (resp SPS5), the decay channel 
$\wt{b}_1 \to W \wt{t}_1$ (resp 
$\wt{t}_2 \to Z \wt{t}_1$) is closed and so the tree level 
width decreases to zero.
The observed peaks around $\cos\theta_t\approx 0.46$ (resp 
$\cos\theta_t\approx 0.53$)
correspond to the opening of the transition 
$\wt{t}_1\to \chi_1^0 t$ (resp $\wt{t}_1\to \chi_2^0 t$).\\
In all cases, we have isolated the 
QED corrections (virtual photons and real photons), we have checked 
that this contribution is very small, less than about 1\%.\\

\begin{figure}[t!]
\smallskip\smallskip 
\vskip-.5cm
\centerline{
{\epsfxsize3.32 in\epsffile{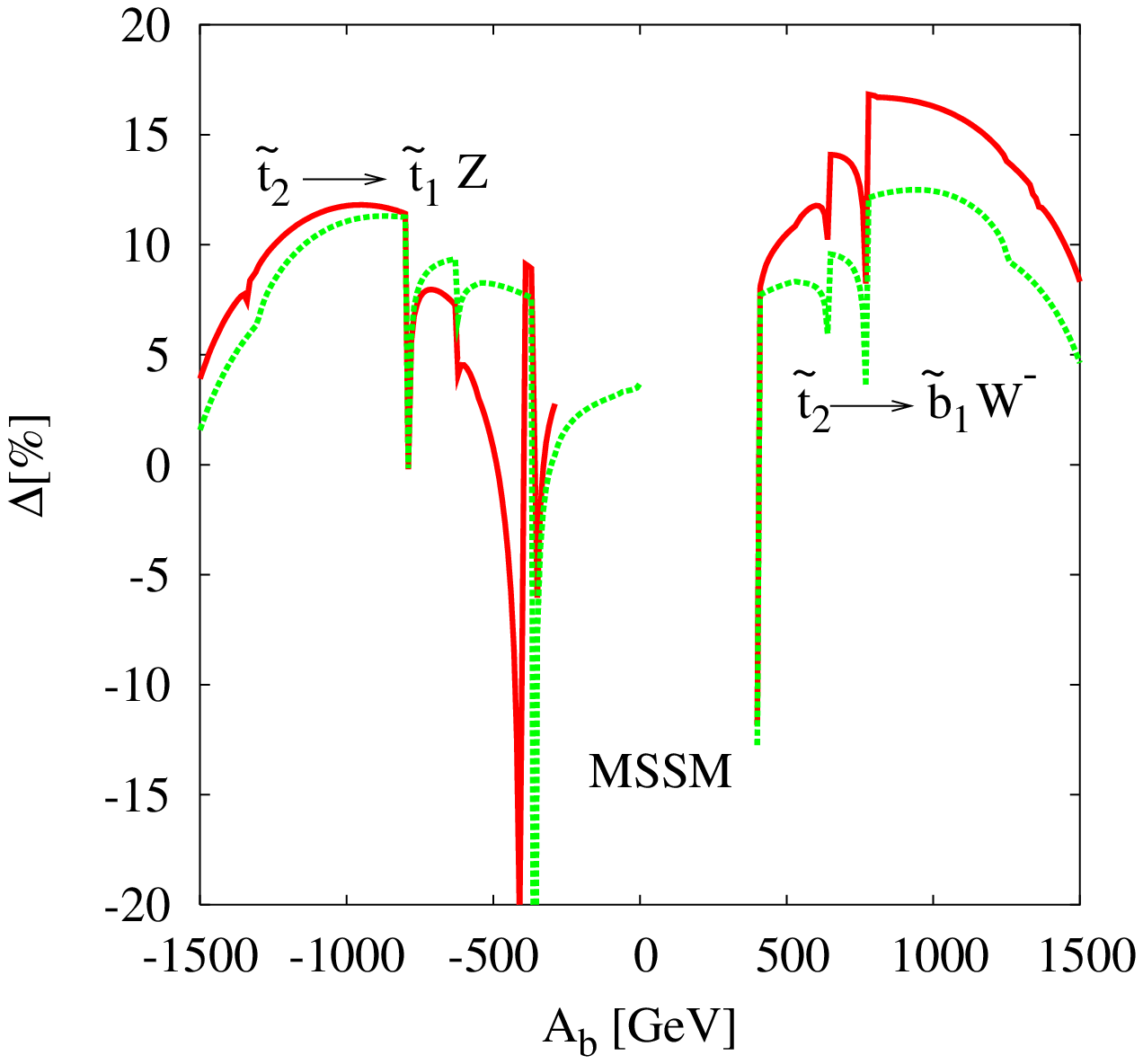}}  
\hskip-0.5cm
{\epsfxsize3.32 in\epsffile{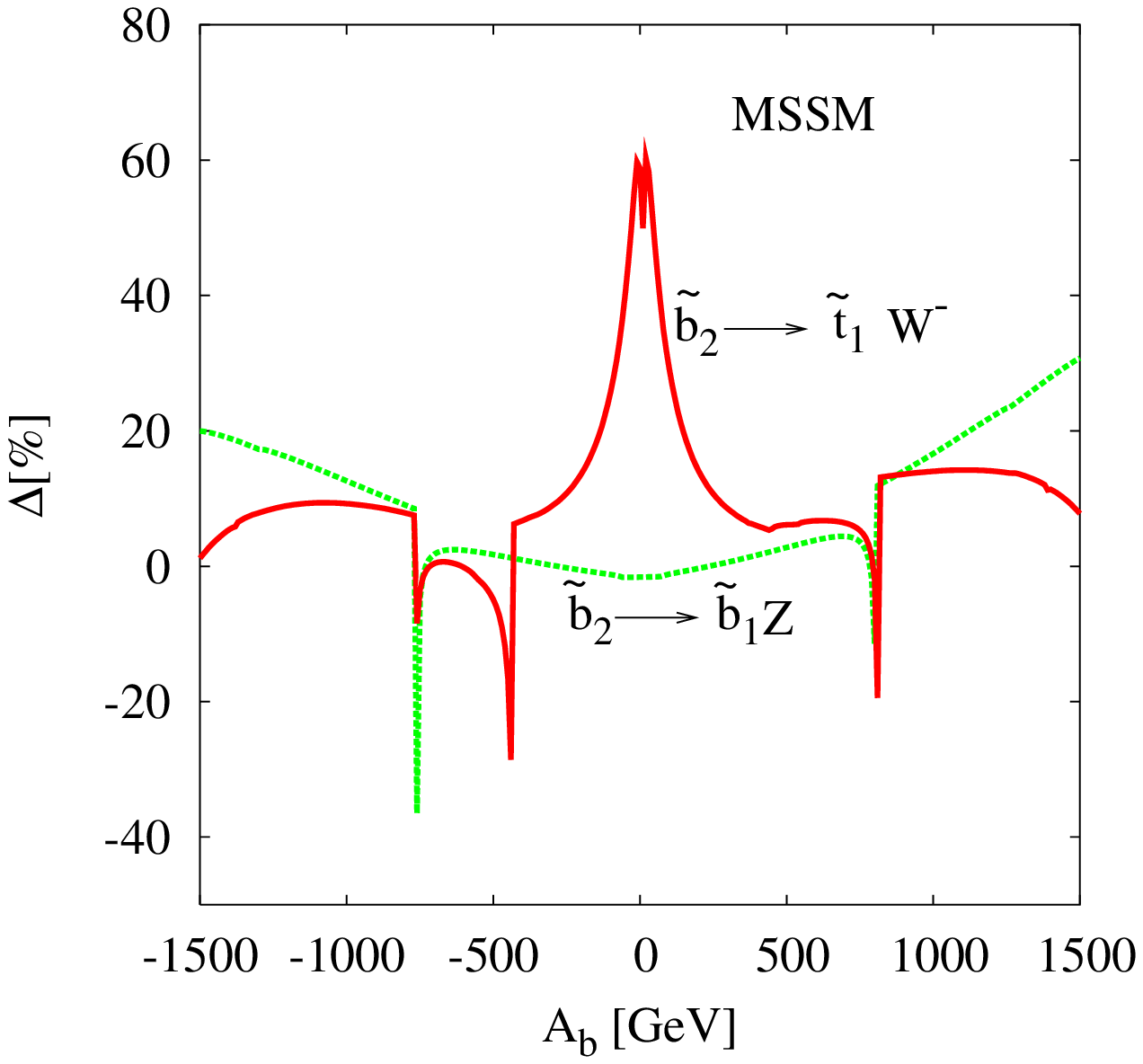}}}
\smallskip\smallskip
\vskip-.58cm
\caption{Relative correction to 
$\wt{t}_2 \to \wt{b}_1 W$,
 $\wt{t}_2 \to \wt{t}_1 Z$ (left)
and 
$\wt{b}_2 \to \wt{b}_1 Z$,
 $\wt{b}_2 \to \wt{t}_1 W$ (right)
as function of $A_t=A_b$ in general MSSM for $\mu=500$ GeV, $M_2=130$ GeV,
 $M_A=200$ GeV and $\tan\beta=60$}
\label{mssm}
\end{figure}

Fig.(\ref{mssm}) illustrates the relative corrections 
to $\wt{t}_2 \to \wt{b}_1 W$, $\wt{t}_2 \to \wt{t}_1 Z$ (left)
and  $\wt{b}_2 \to \wt{b}_1 Z$, $\wt{b}_2 \to \wt{t}_1 W$ (right)
 as function of $A_b=A_t$ in general MSSM for 
large $\tan\beta=60$, $\mu=500$ GeV, $M_2=130$
 GeV and $M_A=200$ GeV.
It is clear from this plot that the relative corrections are bigger 
than in the cases of SPS scenarios.  
This enhancement shows up for large $|A_b|$ and also near threshold regions.
In this scenario, the SUSY-QCD corrections are about 2\%,
the electroweak corrections are about 5\% while the QED corrections
are very small. The dominant contribution comes from the Yukawa
corrections and is enhanced by large $\tan\beta=60$ and large $|A_b|$.\\
In the left plot of Fig.(\ref{mssm}), the region $|A_b|=|A_t|<300$ GeV has no
data. This is due to the fact that splitting between $\wt{t}_2$ and
$\wt{t}_1$ ($\wt{t}_2$ and $\wt{b}_1$) is not large enough to allow 
the decays  $\wt{t}_2 \to \wt{t}_1 Z$
and $\wt{t}_2 \to \wt{b}_1 W$.\\
In the right plot of Fig.(\ref{mssm}), when 
$|A_b|=|A_t|\approx 0$ GeV, the splitting between $\wt{b}_2$ and
$\wt{t}_1$ is close to $m_W$ mass and so the tree level width for
$\wt{b}_2\to \wt{t}_1W^+$ 
almost vanish, consequently the correction is getting bigger.
This behavior has been also observed in previous plots for 
$\wt{b}_2\to \wt{t}_1W^+$.

\begin{figure}[t!]
\smallskip\smallskip 
\vskip-.41cm
\centerline{
\hskip1.1cm
{\epsfxsize3.7 in\epsffile{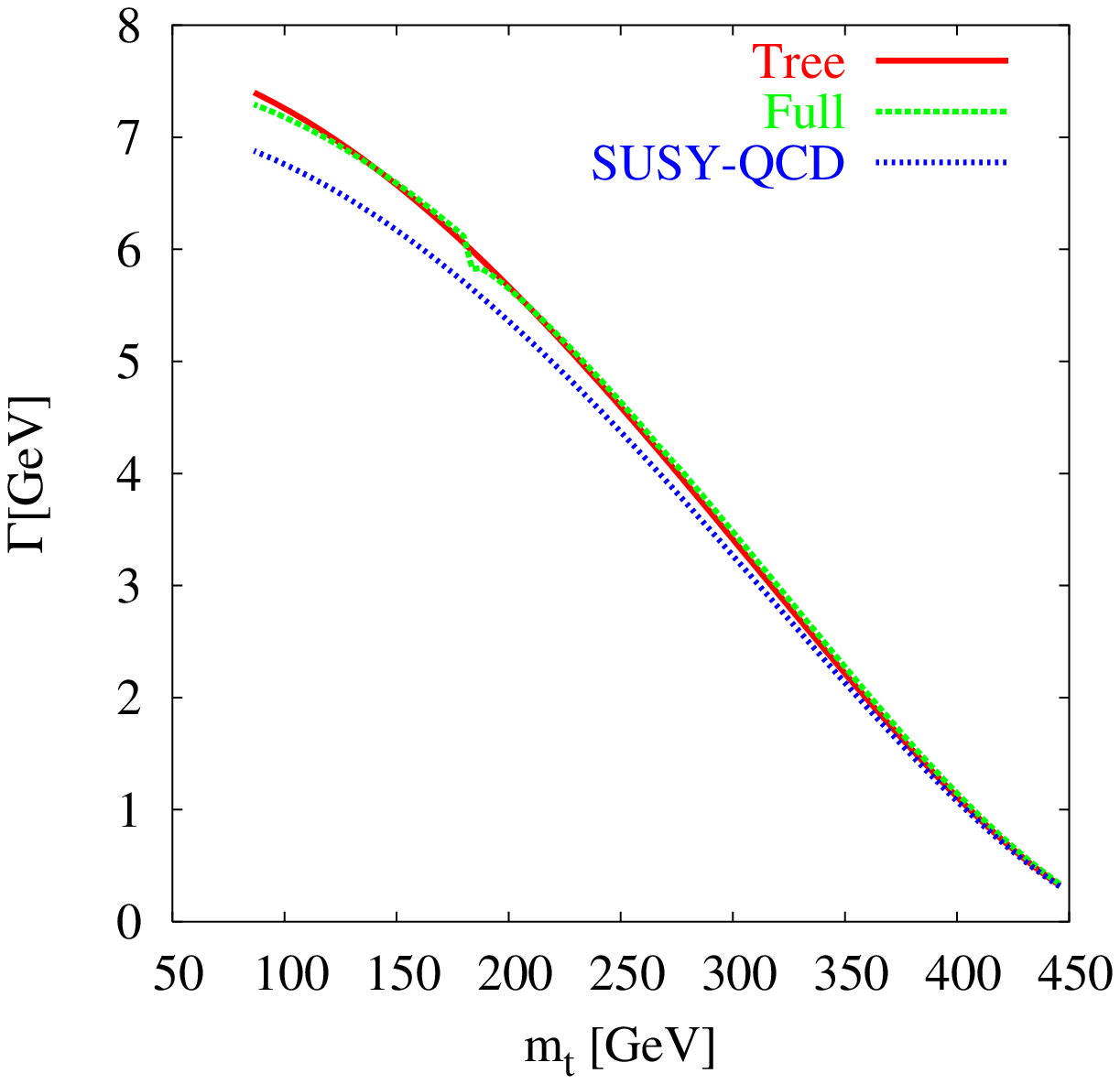}}  
\hskip-1.99cm
{\epsfxsize3.7 in\epsffile{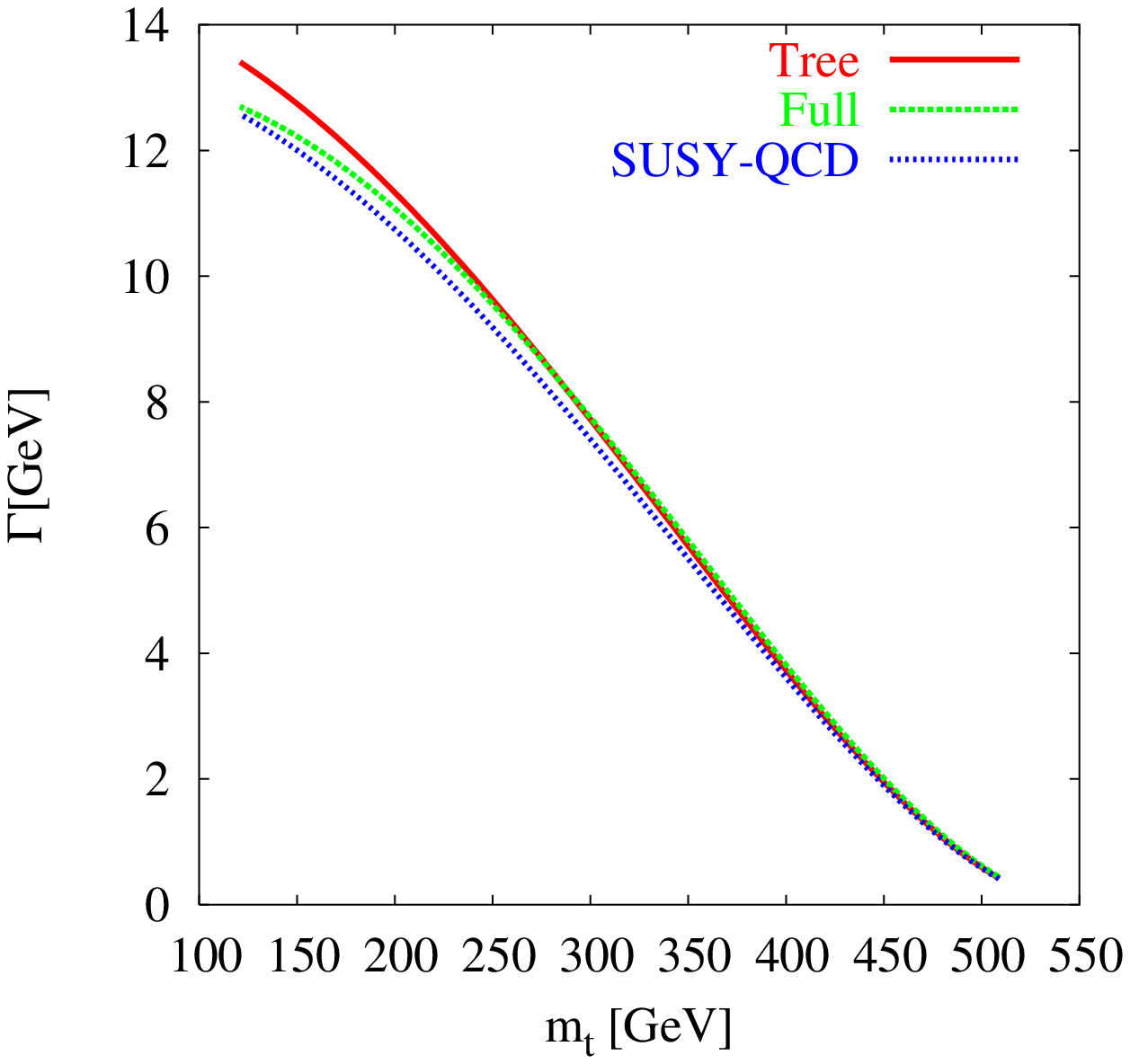}}}
\smallskip\smallskip
\caption{Tree and one loop decay width of $\wt{t}_2 \to \wt{t}_1 Z$ 
as function of $m_{\wt{t}_1}$}
\label{fig6}
\end{figure}

Finally, in Fig.~(\ref{fig6}) we illustrate the decay width of 
$\wt{t}_2 \to \wt{t}_1 Z$ as function of $m_{\wt{t}_1}$ 
in SPS1 (left) and  SPS5 (right). In SPS1
(resp SPS5) the decay width of $\wt{t}_2 \to \wt{t}_1 Z$ 
is about 8 GeV (resp 13 GeV)
for light stop mass of the order 100 GeV. 
Obviously, these decays width decrease 
as the light stop mass increase.

It is clear that the SUSY-QCD corrections reduces the width 
while the electroweak
corrections cancel part of those QCD corrections.
Both in SPS1 and SPS5, the full one loop width of 
$\wt{t}_2 \to \wt{t}_1 Z$  is in some case slightly bigger 
than the tree level width.

\section*{{{Conclusions:}}} 
A full one-loop calculations
of third-generation scalar-fermion decays into gauge bosons W and Z
are presented in the on--shell scheme. 
We include both electroweak, QED and SUSY-QCD
contributions to the decay width. It is found that the QED corrections
are rather small while the electroweak and SUSY-QCD corrections 
interfere destructively.

The size of the one-loop effects are 
typically of the order $-5 \%\to 10$ \% in SPS scenarios which are based on
SUGRA assumptions. While in model independent analysis like 
the general MSSM, the size of the corrections are bigger and can reach 
about 20\% for large $\tan\beta$ and large soft SUSY breaking $A_b$.
Their inclusion in phenomenological studies and analyses are
then well motivated.

\section*{{{Acknowledgment:}}} 
The authors thanks ICTP for the warm hospitality extended to them 
during the first stage of this work.
We thanks Margarete Muhlleitner for useful 
exchange of informations about Sdecay \cite{sdecay2}.
This work is supported by PROTARS-III D16/04.

\end{document}